\documentclass[preprint]{aastex6}

\shorttitle{Linked Scatter Plots}
\shortauthors{Carbon, Henze, and Nelson}

\usepackage{natbib}
\begin{document}


\title{Exploring the SDSS Dataset with Linked Scatter Plots: I. EMP, CEMP, and CV Stars}



\author{Duane F. Carbon, Christopher Henze, and Bron C. Nelson}
\affil{NASA Ames Research Center, NASA Advanced Supercomputing Facility, Moffett Field, CA, 94035-1000, USA;\\ Address correspondence to: Duane.F.Carbon@nasa.gov}

\begin{abstract}

We present the results of a search for extremely metal-poor (EMP), 
carbon-enhanced metal-poor (CEMP), and cataclysmic
variable (CV) stars using a new exploration tool based on linked scatter
plots (LSPs).  Our approach is especially designed to work with very
large spectrum data sets such as the SDSS, LAMOST, RAVE, and Gaia data
sets, and it can be applied to stellar, galaxy, and quasar spectra.  As a
demonstration, we conduct our search using the SDSS DR10 data set.  We
first created a 3326-dimensional phase space containing nearly 2
billion measures of the strengths of over 1600 spectral features in
569,738 SDSS stars.  These measures capture essentially all the
stellar atomic and molecular species visible at the resolution of SDSS
spectra.  We show how LSPs can be used to quickly isolate and examine
interesting portions of this phase space.  To illustrate, we use LSPs
coupled with cuts in selected portions of phase space to extract EMP
stars, CEMP stars, and CV stars.  We present identifications for
59 previously unrecognized candidate EMP stars and 11 previously
unrecognized candidate CEMP stars.  We also call attention to 2
candidate He~II emission CV stars found by the LSP approach that have
not yet been discussed in the literature.

\end{abstract}


\section{Introduction} \label{intro}

With the advent of very large datasets of stellar spectra such as the
SDSS, LAMOST, RAVE, and Gaia (e.g.~\citet{wyse16} for an overview) , it
is no longer remotely possible for an observer to digest the contents
of an entire dataset visually.  The number of spectra is so large that
a dedicated expert team could take years to accomplish even a cursory
examination.  The Henry Draper catalogs (see \citet{henrydraper} and
subsequent volumes) are a well-known and heroic example that
summarized the careful visual examination of ``only'' 225,000 objects.
Because modern datasets can be vastly larger and richer, they are
largely resistant to the methodical examination of spectra by eye
as exemplified by the Draper effort.  For this reason, computer-aided
approaches for extracting useful information from spectrum archives
have become a matter of necessity.

Data-mining of large stellar spectrum archives is generally done with
very specific, predetermined goals in mind. Typical endpoints are
spectral classifications or the estimation of parameters like
$T_{\mathrm{eff}}$, $\log g$, [Fe/H], and [$\alpha$/Fe].  Many approaches have
been used to accomplish these goals; a far less than exhaustive list
includes distance
minimization~\citep{wilhelm99,allende06,covey07,lee08,boeche11,bijaoui12},
projection methods ~\citep{recio06}, decision trees \citep{bijaoui12},
and artificial neural networks
~\citep{bailer-jones00,re_fiorentin07,manteiga10}.  Often,
combinations of these are used in a single data processing pipeline
\citep{lee08,kordopatis13,recio16}.

With all the aforementioned approaches, the researcher must first
choose specific examples of spectra, either observed or synthetic,
which serve to define the characteristics one hopes to extract from
the spectra of the archive.  This set of preselected spectra, referred
to as a ``template'' or ``training'' set, is very problem specific and
can be time-consuming to produce.  However, once the template/training
set is defined and tested, the chosen algorithm can be applied in a
fully automated way to each spectrum in the archive to deduce the
desired classification or stellar parameters.  Large-scale application
of these techniques has led to major advances in our understanding of
galactic chemical evolution by providing very large samples of
abundance and kinematic data for analyses in aggregate
\citep[e.g., ][]{beers08,dierickx10,chiappini15,kordopatis16}. At the same
time, these template/training set techniques have provided a rich pool
of information from which researchers may select individual objects
for more detailed, higher resolution investigations.  These
include numerous studies of low-metallicity stars (examples referenced
in Sections~\ref{emps} and~\ref{cemps}) as well as efforts to select
candidates for such diverse objects as white dwarfs with infalling
accretion disks \citep{gansicke08,wilson14}, Li-rich field giants
\citep{martell13}, and metal poor stars with extreme [$\alpha$/Fe]
excesses or deficiencies \citep{li14,xing14}.

In this paper, we describe an alternative approach to extracting
spectra with interesting desired characteristics from large archives.
We believe it not only has more innate flexibility and nimbleness
than methods using template/training sets but it also provides quicker
feedback on both the success and failings of the user's choices.  To
illustrate, we show how the combination of linked scatter plots and
the NASA Advanced Supercomputing (NAS) hyperwall were used to data-mine a large spectrum dataset,
the Sloan Digital Sky Survey, Data Release 10 (SDSS DR10)
\citep{ahn14}.  In particular, we focus on three subsets of
interesting stellar objects. First are the extremely metal-poor (EMP)
stars whose relevance to the early chemical evolution of our galaxy
has made them the focus of considerable effort in the recent years
(see \citet{frebel15a} for a comprehensive overview).  Next are a subset
of the metal-poor stars, the carbon-enhanced metal-poor (CEMP) stars, which may be
informative of supernova element building in the earliest periods of
galactic chemical enrichment
\citep{beers05,carollo12,beers12,beers14,cooke14,frebel15a}.  Finally,
because the first two examples involved using only absorption features
in our search, we demonstrate how one can additionally use emission
lines to extract and examine objects like the cataclysmic variable
(CV) stars.

Section~\ref{hw_lsp} provides a quick introduction to the NAS hyperwall
and to linked scatter plots (LSPs).  The subsections of Section~\ref{pipeline} describe in
some detail how we populated our working database of nearly 2 billion
feature measurements from the information made available by the SDSS.
We discuss how we determined stellar continuum levels
(Section~\ref{cont}) and how we used these continua to create usable
feature measures (Sections~\ref{feature} and ~\ref{measurement}).  The
reader should be aware that all wavelengths mentioned in this paper
are vacuum wavelengths indicated in
Angstroms~(\AA).  Section~\ref{emps} details how we used our approach
to discover 70 new EMP and CEMP candidates (see
Table~\ref{table_likely_emps} and Table~\ref{table_uncertain_emps}).
Section~\ref{cemps} recounts how one might extract the 11 new CEMP
candidates more directly.  In Section~\ref{cvs} we show how emission
lines can be incorporated in the search as well, in this case, to
discover CV stars. Section~\ref{improvements} describes shortcomings of our
current implementation and outlines needed improvements. Our principal
results are summarized in Section~\ref{summary}.

\subsection{The hyperwall and Linked Scatter Plots} \label{hw_lsp}

The hyperwall at the NASA Advanced Supercomputing Facility, located at
NASA Ames Research Center, is a powerful research tool for the
exploration of large observational and theoretical datasets.  Here we
give a brief introduction to the NAS hyperwall; additional details may be
found in \citet{sandstrom03} and \citet{henze98}.  The hyperwall
hardware is a 23$'$ wide by 10$'$ high wall of 128 workstation displays
arranged in an ($8 \times 16$) array, accompanied by a separate operator's
console with two displays. Each hyperwall display is controlled by a
parallelized software suite running on a node of NASA's Pleiades
supercomputer.  This software is key to the versatility and
exploratory power of the hyperwall.  The operator may choose from a
variety of modes for using the hyperwall, ranging from display of a
complex and active single image involving the whole hyperwall on the
one hand to using each of the 128 displays to portray multiple aspects
of a dataset on the other.  In this study, we employ the latter
capability using linked scatter plots.

Linked scatter plots are the single most critical element of our
approach and a topic that may be unfamiliar to many astronomers.  Here
we briefly describe the experiential aspects of using LSPs on the
hyperwall.  Details of software implementation, which are both highly
complex and site-specific, are sidestepped.

To illustrate our application on the hyperwall, we work with a very
large multivariable data set that includes measured feature
strengths that we derived from stellar spectra of the SDSS DR10.  We
downloaded $N_s$~=~569,738 SDSS spectra and, as we describe below in
Section~\ref{pipeline}, we made 3318 specific feature measurements on
each and every spectrum in the set.  To these feature measurements, we
add seven colors computed from the SDSS \emph{u,g,r,i,z} magnitudes and an
estimate of each spectrum's overall signal-to-noise ratio (S/N) to make a total of
$N_m$~=~3326 measured quantities for each spectrum.  Each of the
$N_m$ quantities is a variable in our multivariable dataset; each
variable contains $N_s$ values of a specific measurement.  To
appreciate the power of LSPs on the hyperwall, it is useful to
consider the ensemble of data as a phase space with $N_m$ dimensions.

We can populate the hyperwall displays however we wish with
2~$\le$~$N_{hw}$~$\le$~128 different two-dimensional scatter plots, each
plot having a unique coordinate pair drawn from the $N_m$
measurements: ($x_i$, $y_j$;~$i$ = 1,2,3,...,$N_{m}$;
$j$ = 1,2,3,...,$N_{m}$; $i \neq j$).  The plots need not share
coordinate variables.  Armed with the hyperwall's large complement of
software tools, the user may choose individual plot scales and
centering, view object distributions projected on $x-$ and $y-$axes, and
select subsets of plotted points on any display.  Subset selection is
accomplished by graphical ``paint brushes'', or by rectangles and 
lines entered either numerically or by user-manipulated graphical
rectangles and triangles.

Because the variables in each of the 128 displays are linked by
the hyperwall software \citep{henze98}, additional powerful,
user-guided interactions are possible.  For example, when we select a
subset of points on any one of the scatter plots (and hence the subset
of the objects represented by those points), the points corresponding
to the selected objects are automatically highlighted in a distinctive
color on all the other scatter plots.  This is possible because the
points in each scatter plot are linked with their corresponding
data objects and the hyperwall software tracks the representation of
those data objects on all the other displayed scatter plots.  The
points on each scatter plot are automatically redrawn when a selection
occurs.  In each instance, the selected points are drawn in the
appropriate color, and on top of the unselected points to prevent
them from being obscured. The hyperwall system is fast enough that
drawing the updates is essentially instantaneous, giving the user
immediate interactive feedback on the effects of a selection.  Our
rectangle selection tool takes particular advantage of this speed.
The user may grab and move the selection rectangle in a particular
scatter plot window and watch in real time as the selected points in
that rectangle are displayed in all the other 127 scatter plot
windows.  This dynamic feature allows the user to discover
relationships between variables that may not have been otherwise
obvious.

The distinctive color aspect of the LSP by itself is very powerful.
It allows one to see exactly how the selected points, which may
represent a distinctive feature in one scatter plot with a particular
pair of variables, are distributed in other scatter plots involving
different variables. This extremely useful capability was also used in
another graphics tool for astrophysics, VIEWPOINTS, which was adapted
from an earlier version of the hyperwall software and developed for
single-display platforms \citep{gazis10}.  Our implementation of the
\emph{selection}$~\Longleftrightarrow~$\emph{distinctive~color}
linkage extends for
an arbitrary number of successive selections.  Thus, having selected a
subset of objects in one of the $N_{hw}$ displays, one may then select
an additional subset of objects in any of the other ($N_{hw}$~-~1)
displays.  The newly selected objects will then be rendered in a
second distinctive color in all of the displays.  \emph{This process
  may be repeated as often as desired, building up ever more complex
  selections}.

In a very powerful addition to the multiple selection capability, the
hyperwall software allows us to treat the selections as logical
Boolean ``AND'' operations.  Thus, for example, we can make a series
of selections in different portions of measurement phase space and
then show on all the displays with a unique color those objects that
simultaneously satisfy all the selection criteria.  The LSP software
provides full checkpoint and restart capabilities so that the user can
safely save and then later restart work involving a series of complex
selections on many displays without loss of effort.

At any time in the successive selection process, the user can view on
a separate large display at the control console the spectra of the all
objects selected up to that time.  The spectrum display window gives
the user full control over the scales of the spectra, marks the
wavelengths of features used in the selections, and allows the user to
scroll rapidly through the spectra, examining and intercomparing them.
This capability is enormously valuable because it gives the user
immediate feedback on how to refine the selection process. The
spectral display window also allows the user to quickly identify spectra of
particular interest, which may be saved for future reference by a
simple mouse click.

We close by pointing out that many of the above features are
  also effectively available in the powerful TOPCAT software package
  \citep{taylor05} developed for single desktop machines.  While
  making successive selections with logical Boolean operations is
  accomplished rather differently under this package, the results
  should be similar in the end to those obtained with LSPs on the
  hyperwall.  We recommend this package for consideration by those
  who want to try LSPs on their dataset but do not have convenient
  access to the NAS hyperwall by reason of geography.

Having now given an overview of the hyperwall and LSPs, we show in the
next section how we selected the $N_s$ spectra and collected the $N_m$
measurements on each spectrum to create our database.  This database,
in turn, enables us to explore the phase space of feature strengths.

\section{POPULATING THE HYPERWALL DATABASE } \label{pipeline}

For the purposes of this paper we use optical spectra drawn from Data
Release 10 of the Sloan Digital Sky Survey.  This choice is for
illustrative purposes only.  Our approach is generally applicable to
any large rich spectrum data set whether of stars, galaxies, or QSOs.
We downloaded as individual FITS files all ``SpecObj'' spectra of DR10
Class~=~STAR and zWarning~=~0 whose median S/N ratio (snMedian)
was $\ge$~4.  This produced a spectrum dataset containing reasonable-quality spectra for 569,738 unique stellar targets.  For each of the
targets, we also also downloaded such associated quantities as the
target's R.A., decl., $z$, zErr, snMedian, and \emph{u,g,r,i,z} psfMag
magnitudes.  The DR10 data contain spectra from both the SDSS and the
BOSS spectrographs, the latter covering a somewhat wider spectral
range.  The spectrum data includes the vacuum wavelength, calibrated
flux ($10^{-17}$~erg~cm$^{-2}$~s$^{-1}$~\AA$^{-1}$), and
pixel-by-pixel inverse-variance values for each object.  In order to
derive the final dataset that we use on the hyperwall, each spectrum
must be processed through a pipeline coded using
MATLAB\copyright~\citep{matlab11} whose principal steps we describe
next.  Before the first pipeline step, all spectrum wavelengths were
corrected to zero rest velocity using the value of redshift, $z$.

\subsection{Pipeline I --- Continuum Determination} \label{cont}

In order to make the spectral feature measurements used for the linked
scatter plots on the hyperwall, we first must determine a practical
reference level, a ``continuum'', in each spectrum against which
measurements are made.  We note that this continuum is not the true
bound-free/free-free continuum of a stellar atmosphere.  Rather, it is
an operationally chosen smooth practical background whose level, for
hotter or very metal poor stars, might approach the true continuum but
which generally lies well below it.  The degree of deviation of the
observed continuum from the true continuum depends on the presence of
atomic and molecular lines and the spectral resolution.  We need an
automated way of placing continua deemed acceptable to a human
observer over the very large range of spectral types and abundance
variations present in the SDSS DR10 dataset.  Moreover, the procedure
must be capable of handling the breaks in spectral data that are a
not uncommon feature of many DR10 spectra. Our approach largely
satisfies these requirements.

\begin{figure}[t]
\epsscale{0.8}
\plotone{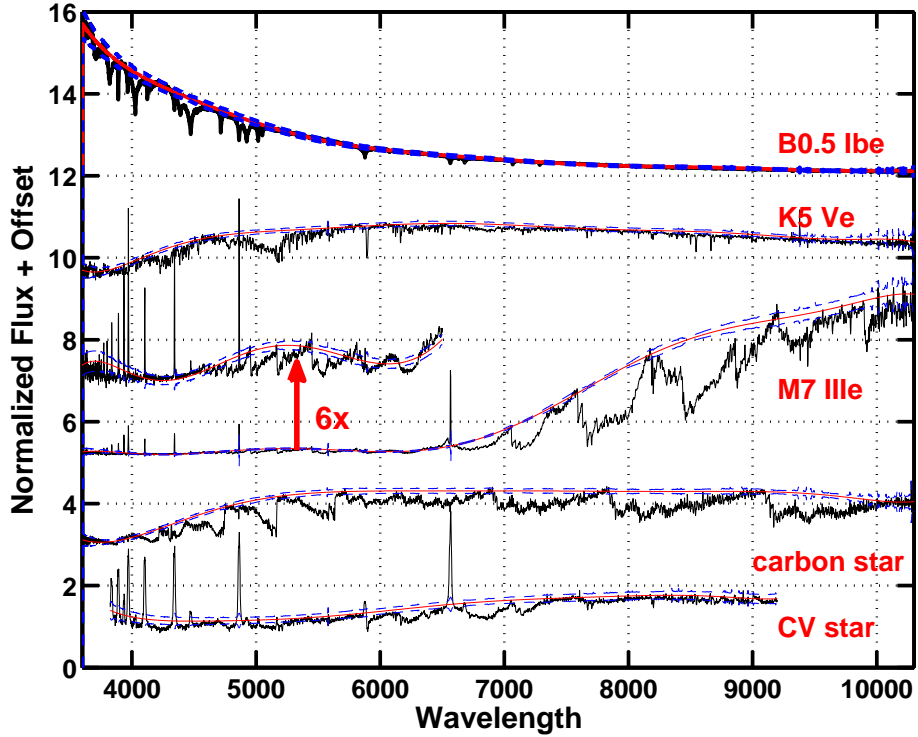}
\caption{Sample automated continuum fitting results with 
  spectral types (SDSS ``subClass'') noted in red. The automatically fit
  continua are shown in red; the blue dashed lines show
  $+/-$~2.5~$\sigma$ about the continuum computed from the
  pixel-by-pixel variance.  For the sake of plot clarity, the
  occasional very large values of variance in the M7~IIIe spectrum
  have been suppressed.  The low-signal short-wavelength end of the
  M7~IIIe star is offset replotted at 6x for easier examination.  Each
  spectrum is normalized by its average continuum flux and then
  shifted to avoid overlap. \label{fig1}}
\end{figure}

We give here an overview of the procedure we employed to determine the
continuum for each spectrum in our downloaded DR10 dataset. A detailed
description will be given elsewhere, and the code will be 
provided there as open source.  In our procedure we first
break the spectrum into intervals of roughly 200~\AA \ width, a number
that was chosen empirically so as to work best with the broad
molecular absorption features of DR10 cool star spectra.  Within each
of these intervals, we apply an iterative algorithm that effectively
eliminates those wavelengths showing emission significantly above the
local noise level.  The remaining wavelengths with the highest fluxes
are then used to define the continuum level for the interval. Once all
intervals are processed, the collection of interval continuum points
are fitted with a ninth-degree polynomial.  If any of the fitting
points fall more than two flux standard deviations below the
polynomial fit, those fitting points are dropped and the ninth degree
polynomial fit is redone with the remaining points.  This last step is
very effective in preventing the continuum from following large
molecular features at the SDSS/BOSS resolutions.  The final fit is our
adopted continuum for each spectrum.

In Figure~\ref{fig1} we show some typical examples of our automated
continuum fits. For clarity the selected objects all have high
S/N spectra, and they represent a variety of objects posing
different challenges to the continuum fitting.  A quick examination
shows that the deduced continuum nicely follows the overall contours
of the spectrum while accurately preserving the innate character of
emission and absorption features.  This is true even when the spectrum
contains mixed atomic emission and broad molecular absorption
features.  A closer look, however, reveals some disappointing flaws.
For example, because we have used a polynomial fit for the final
continuum, the continuum level can go awry at the very ends of the
spectra, particularly if those spectral regions are comparatively
noisy. The continua at the very blue end of the M7 IIIe and CV star
spectra are examples of this behavior; in both cases, the continuum is
placed higher than one would want for a best fit.  Moreover, a close
inspection of the detailed fits in Figure~\ref{fig1} reveals interior
sections where one might have drawn the continuum differently if
working by eye.  However, since such an undertaking for the whole
DR10 dataset is quite impossible, we have used our procedure to
automatically determine the required 569,738 independent continua.  We
give an example in Section~\ref{cvs} of the type of problem that may
created if the continuum fitting is not done well.

\subsection{Pipeline II --- Selecting the Spectral Features to Measure} \label{feature}

In the next section, \ref{measurement}, we describe how we measured the
spectral feature strengths that make up our hyperwall database. Here
we describe how we selected which features to measure in the first
place.  Since all the objects in our data set are (or at least
believed to be) stellar, we aimed to include all principal spectral
features (atomic, ionic, and molecular) that might conceivably be
measurable at the SDSS/BOSS resolution in some DR10 stars.  For the
atoms and first ions of all elements which are even moderately
abundant in the sun, we downloaded all the parameters necessary to
compute individual line absorption strengths from the NIST online
tabulation \citep{nist14}.  For each line, we computed the value of
\begin{equation}
\lambda^2 \, g_{u}\, A_{u,l}\, e^{(-1.44*\omega_{l}/T)}
\end{equation}
which is proportional to the transition's absorption cross section, where
$\lambda$ is the transition wavelength, $g_{u}$ is the upper state
statistical weight, $A_{u,l}$ is the Einstein A value, $\omega_{l}$
is the excitation of the transition's lower state in wavenumbers,
and $T$ is the temperature in kelvins.  
The calculation was done at two
temperatures: 8200~K and 2895~K; the first is roughly appropriate for
the line-forming region in the photosphere of a $T_{\mathrm{eff}} = 10000~K$
star and the latter for the line-forming region of a
$T_{\mathrm{eff}} = 3640~K$ star.  These $T_{\mathrm{eff}}$ correspond to an A0 star
\citep{theo91} and an M3 III star \citep{perrin98}, respectively.  For
the most abundant elemental species, we selected for inclusion in our
feature set all atomic and ionic NIST transitions within the BOSS
wavelength range that were at least 1/100 as strong as the strongest
transitions at either of the two selected temperatures.  For less
abundant species, we chose only the strongest lines.  To this
NIST-based feature set, we added all the atomic and ionic emission
lines reported in low-resolution spectrum studies by \citet{merrill47},
\citet{sanford49}, and \citet{cohen80} as well as selected additional lines from
\citet{gray09}.  We added all the principal band head wavelengths
culled from the literature for the main isotopologues of the following
molecules: AlO, ScO, TcO, VO, YO, ZrO, AlH, CaH, CrH, FeH, MgH, SiH,
TiH, C$_{3}$, SiC$_{2}$, and TiS.  Finally, we added the principal
band heads from the following isotopologues: $^{12}$C$^{12}$C,
$^{12}$C$^{13}$C, $^{13}$C$^{13}$C, $^{12}$CH, $^{13}$CH,
$^{12}$C$^{14}$N, $^{13}$C$^{14}$N, $^{12}$C$^{15}$N,
$^{47}$Ti$^{16}$O, $^{48}$Ti$^{16}$O, $^{49}$Ti$^{16}$O, and
$^{50}$Ti$^{16}$O.  The final feature list contained 1659 wavelengths,~$\lambda_i$:
926 atomic and ionic and 733 molecular.

\subsection{Pipeline III --- Measuring the Features} \label{measurement}

We adopted the feature strength, $S(\lambda_i)$, as our measure of the
amount of absorption or emission at wavelength~$\lambda_i$:
\begin{equation}
S(\lambda_i) = \frac{[F_c(\lambda_i) - F_l(\lambda_i)]}{F_c(\lambda_i)}
\label{def-S}
\end{equation}
where $F_c(\lambda_i)$ is the continuum flux and $F_l(\lambda_i)$ is
the observed spectrum's flux at feature wavelength $\lambda_i$.  We note that for a fully
absorbed spectral line $S(\lambda_i)$~$\rightarrow$~1.0; emission lines
may have arbitrarily large negative values of $S(\lambda_i)$.
We have also adopted a second feature strength measure,
$D(\lambda_i)$, which we have often found helpful when searching for
solid detections of weak features:
\begin{equation}
D(\lambda_i) = \left|\frac{[F_c(\lambda_i) - F_l(\lambda_i)]}{\sigma(\lambda_i)}\right|
\end{equation}
where $\sigma(\lambda_i)$ is an estimate of the noise level at $\lambda_i$ determined
from the interpolated pixel-by-pixel inverse-variance for each
spectrum. $D(\lambda_i)$ helps to cull out those objects for which the
apparent feature is likely to be just noise.  $S(\lambda_i)$ and
$D(\lambda_i)$ were measured at each of the 1659 feature wavelengths
for each of the 569,738 spectra, generating nearly 2 billion
measurements from our SDSS DR10 dataset.  For clarity in the following
text, we will refer to $S(\lambda_i)$ and $D(\lambda_i)$ by a
feature's identification rather than its wavelength,
e.g.,~$S(\mathrm{H}_\alpha)$ or $S(\mathrm{Ca~II~K})$.  When there may be
ambiguity, we will indicate the wavelength, e.g.,~$S(\mathrm{CH~4308})$.

In closing this section we wish to stress the simple
point that carrying out feature measurements at specific wavelengths
does \emph{not} mean that one is actually measuring the desired species in
all the SDSS stellar spectra. Because the SDSS database contains
within itself representatives of nearly all stellar types from hot to
cold, oxygen-rich to carbon-rich, considerable confusion can be
expected especially given the limited spectral resolution.  For
example, a measure for He I line absorption at 3889.75~\AA,
appropriate for hot stars, will be dominated in cooler stars by the
blue-degraded $^{12}$CH~B-X~(0,0) absorption at 3890.10~\AA.  There
are numerous opportunities for such ambiguity, and appropriate care
must be taken in interpreting phase-space selections like those that
we describe in the next section.

\subsection{Why Uncalibrated Feature Strengths? } \label{uncalibrated}

We wish to emphasize that we have made no effort to calibrate the
feature measures described above so that they might yield derived
quantities such as [$\alpha$/Fe], or any elemental abundances.  While
possible, such an undertaking would be complex and difficult and
require an extensive grid of synthetic spectra covering the entire
range of spectral types represented in the SDSS archive.  Researchers
referred to in Section~\ref{intro} have undertaken such efforts for
limited cases with considerable success.  Here, on the other hand, we
work with the simplest of feature measures, line depths, to explore a
very large and rich archive of spectrum data.  Despite this
limitation, we will show in Sections~\ref{emps}, \ref{cemps}, and
\ref{cvs}, that we can quickly isolate interesting objects worthy of
further study.

Because our database contains feature measures for nearly every atomic
feature and molecular band likely to be encountered in the SDSS
spectra, we have great flexibility in deciding which combinations of
feature measures and selections we might like to examine in any
hyperwall session.  We are not tied to the limitations imposed by
working with a narrow set of predefined and carefully calibrated line
indices.  Instead we have the full range of spectral features visible
in the SDSS stellar spectra at our disposal.  This provides great
freedom for parsing the content of the SDSS dataset and encourages
experimentation and exploration.  It not only allows us to query the
SDSS spectra for the presence of spectra whose characteristics were
not anticipated when our pipeline was set up, but also opens the
possibility for serendipitous discoveries.

That said, there is nothing to prevent future versions of our pipeline
from adding useful, established indices in addition to the measures
now made on each spectrum.  In Section~\ref{improvements}, we indicate
which line indices will be included in the next version of the
pipeline.

\section{APPLICATIONS OF THE HYPERWALL AND LSPs } \label{applications}

In the following subsections, we will discuss three straightforward
applications of the tools we have described in an effort to make the
hyperwall and LSP capabilities more concrete for potential users.  In
the first application, we show how LSPs can be used to isolate objects
in the DR10 dataset that are candidate EMP and CEMP stars. Next, we
will show how one can directly locate candidate CEMP stars.  Finally,
we will show how one can isolate DR10 objects which are candidate CV
stars with He~II emission.  In each case we make simple selections in
some small set of phase spaces to isolate candidate objects of
interest.  \emph{We emphasize that the selections we make in the
  following sections are meant only to illustrate the LSP approach. The
  parameters of our choices are not intended to be definitive.}
Experts deeply familiar with the spectra of the objects we seek could
easily be expected to make different choices as well as add cuts in
other 2D phase spaces beyond the small set we use. We welcome input
from interested researchers.

\subsection{Application I --- EMP Stars} \label{emps}

\subsubsection{Selecting by S/N and  $u$-$z$} \label{selection1}
We begin by illustrating how one might use LSPs to identify EMP
stars.\footnote{We adopt the definition of \citet[][
    Table~I]{beers05}} While many approaches to this problem based on
feature strengths might be devised, the one we use here is
deliberately simple.  To get an idea of the phase space parameters
that might isolate EMP stars, we examined the colors and line
strengths for a sample of 168 known SDSS EMP stars discussed in the
literature
\citep{aoki08,lai09,behara10,bonifacio11,caffau11b,caffau11a,bonifacio12,caffau12,carollo12,sbordone12,aoki13,caffau13,lee13,caffau14,bonifacio15,frebel15b,placco15}.

\begin{figure}
\epsscale{0.5}
\plotone{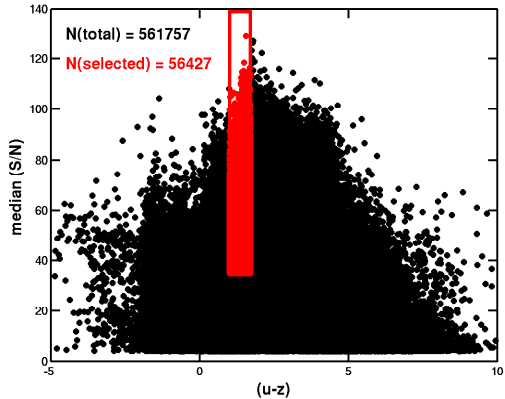}
\caption{Culling by ($u$-$z$) and S/N.  All the downloaded stars within the 
figure boundaries are shown in black.  The selection (in red) extracts the 
subset of those stars that satisfy the constraints
(1.0~$\le$~($u$-$z$)~$\le$~1.7) and (median~S/N~$>$~35).  The color-coded 
``N''s in the figure indicate how many stars are in the figure area and how
 many are subselected at this stage. \label{fig2_u_z_sn}}
\end{figure}

This review led to the following cuts in feature measurement
phase space. Our first cuts were made in the ($u-z$, median~S/N)
plane, see Figure~\ref{fig2_u_z_sn}.  We chose stars whose median~S/N
was greater than 35 since it becomes increasingly difficult to make clear
judgements as to metallicity when the median~S/N falls below that value; the
cut in $u$-$z$ color is guided by the $u$-$z$ range of the 168 known
SDSS EMP stars.  This first cut selects roughly 10\% of the stars in the
whole dataset.

\subsubsection{Subselecting by $S(\mathrm{H}_\epsilon)$ and $S(\mathrm{Ca~II~K})$} \label{selection2}

\begin{figure}
\epsscale{0.5}
\plotone{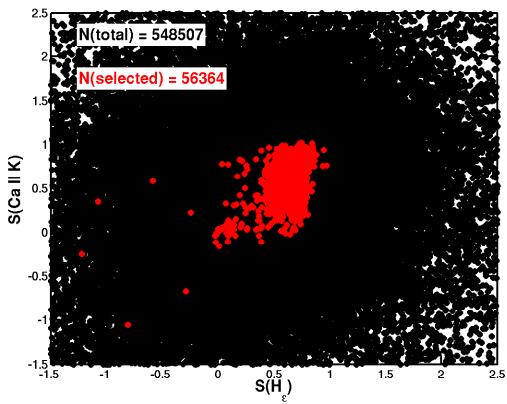}
\caption{ Distribution in the ($S(\mathrm{H}_\epsilon)$, $S(\mathrm{Ca~II~K})$) plane of 
stars previously selected in the ($u$-$z$, median~S/N) plane.  All the 
downloaded SDSS stars within the figure boundaries are shown in black.  
Those selected by the cuts ((1.0~$\le$~($u$-$z$)~$\le$~1.7), 
(median~S/N~$>$~35)) in the ($u$-$z$, median~S/N) plane are shown in red.  
The color-coded ``N''s indicate how many stars are in the figure area 
and how many in the area were previously selected.\label{fig3_HepsCaIIK_all}}
\end{figure}

Next, we turn our attention to the ($S(\mathrm{H}_\epsilon)$, $S(\mathrm{Ca~II~K})$)
plane.  It is instructive to first examine this portion of feature
strength phase space with the stars selected using the ($u$-$z$,
median~S/N) cut superimposed over all the stars in the relevant
portion of the ($S(\mathrm{H}_\epsilon)$, $S(\mathrm{Ca~II~K})$) plane,
Figure~\ref{fig3_HepsCaIIK_all}.

In addition to the general placement of our selected subset in the
($S(\mathrm{H}_\epsilon)$, $S(\mathrm{Ca~II~K})$) plane, two particular points should be
noted.  First, given the values of N in this figure and in
Figure~\ref{fig2_u_z_sn}, not all stars selected by the ($u$-$z$,
median~S/N) cut lie within the bounds of the current figure; if
desired, this could be cured by expanding the plot domain.  Second,
clearly many of the objects in Figure~\ref{fig3_HepsCaIIK_all} have
$S({\lambda}_i)$~$>$~1.0, fully 8.4\% of the population seen in this
figure.  Naively, one might have expected such values to be an
impossibility given the definition of $S(\lambda_i)$.  These aberrant
values occur because when the signal level in a SDSS spectrum is
small, negative values of flux often occur.  These, when combined with
positive values for the continuum flux consistently, yield
$S(\lambda_i)$~$>$~1.0.  This effect is particularly exacerbated when
$F_c(\lambda_i)$ approaches zero in the denominator of
Equation~\ref{def-S}.  It should be noted that the stars that satisfy
the median~S/N~$>$~35 constraint do not show these anomalous values
since higher median~S/N generally means higher signal strength in the
blue.

\begin{figure}
\epsscale{0.5}
\plotone{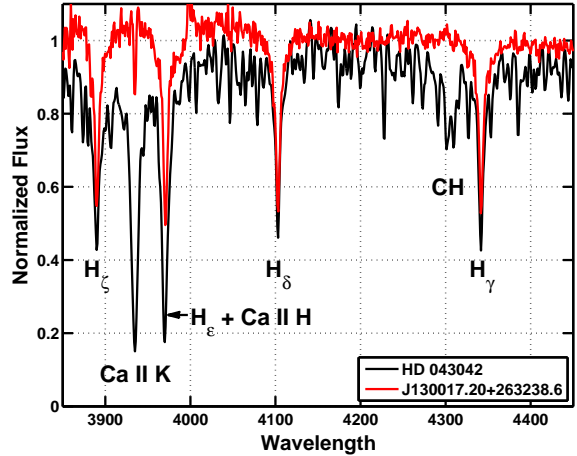}
\caption{Comparison of the SDSS spectrum of a known EMP star (J130017.20+263238.6 in red) and a solar abundance star (HD~043042 in black).  The stars have essentially the same $T_{\mathrm{eff}}$ and $\log g$. Each spectrum is normalized by its average flux over the interval [4194,4195].  Atomic and molecular features prominent in EMP or CEMP spectra at the SDSS spectral resolution are specifically indicated.   \label{fig4_nonemp_emp}}
\end{figure}

\begin{figure}
\epsscale{0.5}
\plotone{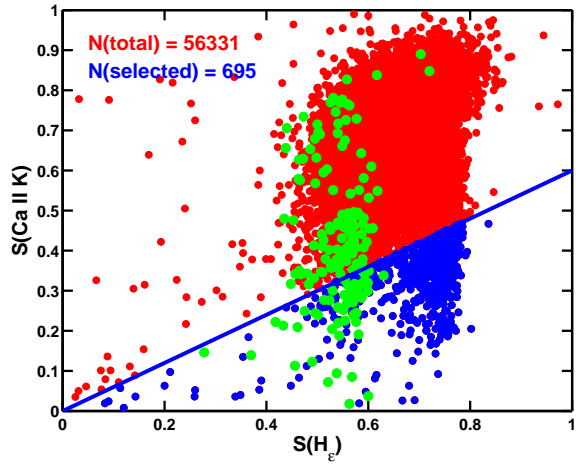}
\caption{Culling potential EMP candidates.  All the red and blue stars 
in the figure satisfy the constraints: (1.0~$\le$~($u$-$z$)~$\le$~1.7) 
and (median~S/N~$>$~35).  The stars indicated in blue \emph{additionally} 
satisfy the constraint that ($S(\mathrm{Ca~II~K})$~$<$~0.6~$\times$~$S(\mathrm{H}_\epsilon)$), 
a restriction that makes the selected stars likely EMP candidates.  
The color-coded ``N''s indicate how many stars are in the figure area 
and how many were subselected by the diagonal cut. For reference, the 
166 known SDSS EMP stars that have both $S(\mathrm{Ca~II~K})$ and $S(\mathrm{H}_\epsilon)$ 
measures are shown in green. \label{fig5_HepsCaIIK_partial}}
\end{figure}

 Now, examining the known SDSS EMP stars, we look for parameters that
 will isolate the stars that are most likely to be EMP candidates from the
 general population.  Figure~\ref{fig4_nonemp_emp} compares the SDSS
 spectrum of a extreme EMP star (J130017.20+263238.6 (6393, 4.00,
 [M/H] = -3.65) \citet{bonifacio12}) with a near solar metallicity
 star (HD~043042 (6392, 4.22, [M/H] = 0.08) \citet{sanchez06}) from
 the MILES Library.  Since the effective temperatures and surface
 gravities of the two stars agree so closely, differences in
 appearance are due to metallicity alone.  The dramatic weakening of
 the metal lines in the EMP star is clearly evident.  It is
 immediately obvious that the signature low ratio of
 $S(\mathrm{Ca~II~K})$/$S(\mathrm{H}_\epsilon)$\footnote{Note that because of the
   limited SDSS resolution, the depth of $\mathrm{H}_\epsilon$ is impacted by the presence of the unresolved, overlapping $\mathrm{Ca~II~H}$ line.  Nevertheless, $\mathrm{H}_\epsilon$ clearly dominates for the stars we are interested in.} might be a useful selection constraint.  For
 J130017.20+263238.6 this ratio is 0.39, whereas the ratio is
 $\approx$~1.0 for HD~043042.  While many EMP stars have $S(\mathrm{Ca~II~K})$
 comparable to, and even larger, than their $S(\mathrm{H}_\epsilon)$, stars in
 the ($S(\mathrm{H}_\epsilon)$, $S(\mathrm{Ca~II~K})$) plane that satisfy the constraint
 $S(\mathrm{Ca~II~K})$~$<$~0.6~$\times$~$S(\mathrm{H}_\epsilon)$ are very likely to be
 EMP stars.  This cut is shown in Figure~\ref{fig5_HepsCaIIK_partial}
 where we have restricted ourselves to the relevant absorption portion
 (0.0~$\le$~$S(\lambda_i)$~$\le$~1.0) of the
 ($S(\mathrm{H}_\epsilon)$,~$S(\mathrm{Ca~II~K})$) plane and show only those stars that
 already satisfy the ($u$-$z$,~median~S/N) cut.  For reference, we
 also show in Figure~\ref{fig5_HepsCaIIK_partial} the positions of the
 known SDSS EMP stars mentioned above.  Close examination reveals that
 only 33\% of the known EMP stars lie within the currently selected
 region.  Fully two-thirds of known SDSS EMP stars have
 $S(\mathrm{Ca~II~K})$~$\geq$~0.6~$\times$~$S(\mathrm{H}_\epsilon)$; indeed, 23\% lie above
 the line defined by $S(\mathrm{Ca~II~K})$~$=$~$S(\mathrm{H}_\epsilon)$.  Thus, while
 those currently selected are excellent candidates, it is clear that
 different constraints would add to the pool of candidates.  We chose
 not to broaden the search area primarily because with our current
 constrained variables, $u$-$z$, median~S/N, $S(\mathrm{Ca~II~K})$, and
 $S(\mathrm{H}_\epsilon)$, it was not clear to us how to disambiguate EMP stars
 from stars with higher metallicity on the basis of our current
 measurement set.  Perhaps a cleverer set of constraints than the ones
 we have chosen would be able to net a larger sample of high-likelihood EMP candidates.  Again we encourage suggestions from our
 readers.

\subsubsection{Eliminating obvious false positives - step 1} \label{culling1}

The described cuts in the ($u$-$z$, median~S/N) and ($S(\mathrm{H}_\epsilon)$,
$S(\mathrm{Ca~II~K})$) phase spaces isolated 695 EMP candidates (0.12\% of all
the stars), whose spectra were then examined visually in two steps.
The first step requires only a brief visual examination of each
spectrum and is accomplished quite quickly.  This step is crucial in
weeding out two groups of stars: stars with badly flawed SDSS spectra
at H$_\epsilon$ and/or Ca~II~K, and stars with good spectra that
matched the selection criteria, but which were obviously not classical
EMP stars.  The latter were either early type stars (late-B, early-A),
distinguished by their broad H line wings and steep SEDs, or white
dwarf + cool star binaries/superpositions.  After this first culling, 139
stars remained as apparently possible EMP candidates.  It is important
to note that 30 of these 139 have previously been established as EMP
stars through model atmosphere abundance analysis; 21 by
\citet{aoki13}, and the rest by \citet{caffau11a}, \citet{caffau11b}, \citet{caffau12},
\citet{bonifacio12}, \citet{sbordone12}, \citet{caffau13} and
\citet{li14}. \emph{By itself, this is a strong independent
  confirmation that careful selections even in the limited
  4-dimensional phase space we have considered can have real discovery
  power.}  We find it particularly satisfying that one of the known EMP stars captured by our selections is J102915.15+172928.0, currently the most metal-deficient SDSS EMP star known \citep{caffau11b,caffau12}.

\subsubsection{Eliminating more subtle false positives - step 2} \label{culling2}

Since a detailed model atmosphere abundance analysis of these objects
is well outside the scope of this paper, we attempted to decide which
of the remaining 109 (corrected for the 30 known EMP stars) stars were
the most likely EMP candidates by systematically comparing their SDSS
spectral energy distributions (SEDs) to those of the 168 known SDSS EMP
stars.  The set of known SDSS EMP stars is not without shortcomings,
unfortunately.  First, the stars in this set often have SDSS spectra
that are quite noisy or have poor correction for terrestrial
absorption/emission.  These defects often make them poor comparison
stars.  Second, while classified as EMP stars by the authors of the
papers making the abundance analyses, their metallicities do not
necessarily adhere to the metallicity class definitions laid down by
\citet[][ Table~1]{beers05} and now widely adopted.  A
significant number of stars in the sample of 168 known SDSS EMP stars
are therefore probably more likely on the metal-poorer end of Beers and Christlieb's
``very metal poor'' classification.  However, given the often large
differences in metallicity (as well as in $T_{\mathrm{eff}}$ and $\log~g$) that
one encounters for the same star as analyzed by different
investigators, it is difficult to be too strict in assigning a star to
a particular metallicity classification based on the current
literature.  As a consequence, we adopted the attitude that if a star
has the spectral characteristics, as defined below, of stars in the
sample of 168 known SDSS EMP stars, we shall select it as a ``likely''
EMP candidate star.  Those appropriately armed with high-resolution
spectra and careful model atmosphere abundance analyses will have the
final word.

The spectrum of each of the 109 stars in the remaining candidate list
was compared with the spectra of known SDSS EMP stars.  This somewhat
time-consuming task was made considerably more tractable by using the
SCAMP spectrum visualization tool developed here \citep{idehara04}.
Particular attention was paid to the fit of the known and candidate
SEDs over the entire spectral range.  This helped us weed out luminous
A stars that have $S(\mathrm{Ca~II~K})$/$S(\mathrm{H}_\epsilon)$ ratios in our selected
domain but significantly steeper SEDs than the known EMP stars.  For
objects with closely fitting SEDs, careful attention was then given to
how well the depths and widths of the H Balmer lines agreed between
the candidate and known EMP stars.  We also affirmed that any other
metal lines appearing in the spectrum were consistent with the
strengths in similar known SDSS EMP stars.  If the overall SED and
Balmer lines fits were close, other metal lines appropriately weak or
absent, and the Ca~II~K line in the candidate spectrum was close to or
weaker than those in the spectra of (generally) two or more matching
known SDSS EMP stars, we selected the candidate star as a ``likely''
EMP candidate.  Occasionally, we had to relax these criteria slightly,
usually because the closest matching known SDSS spectra were too noisy
for a clear determination.

\subsubsection{The likely EMP and CEMP stars} \label{results_emps_cemps}

Table~\ref{table_likely_emps} identifies the 57 candidate EMP stars
chosen for their similarity to known EMP stars.  Along with their
standard SDSS star name, we include the plate, fiber, and modified
Julian dates (MJD) of the observed spectrum we used, the $g$
magnitude, the $u$-$g$ and $g$-$r$ colors, relevant $S(\lambda_i)$ and
$D(\lambda_i)$ values, the spectral type (SpT) assigned by the SDSS
pipeline (subClass), plus any additional comments regarding the
star's spectrum.  The apparent presence of the $^{12}$CH A-X (0,0)
band is explicitly noted.  It should be observed that the most
frequently occurring spectral type in Table~\ref{table_likely_emps} is
A0 with 42 stars (74\%) so assigned.  This is not much different
than the result for the 168 known SDSS EMP stars where 66\% of the
stars were assigned the A0 spectral type by the SDSS.  This
suggests that the A0 stars in the SDSS dataset could be a fertile
search area for additional EMP and CEMP stars provided one uses the
right search constraints.  Since there are 56,661 stars classified as
A0 in our downloaded dataset, the importance of good constraints
for EMP/CEMP subselection is clear.

\begin{figure}
\epsscale{0.75}
\plotone{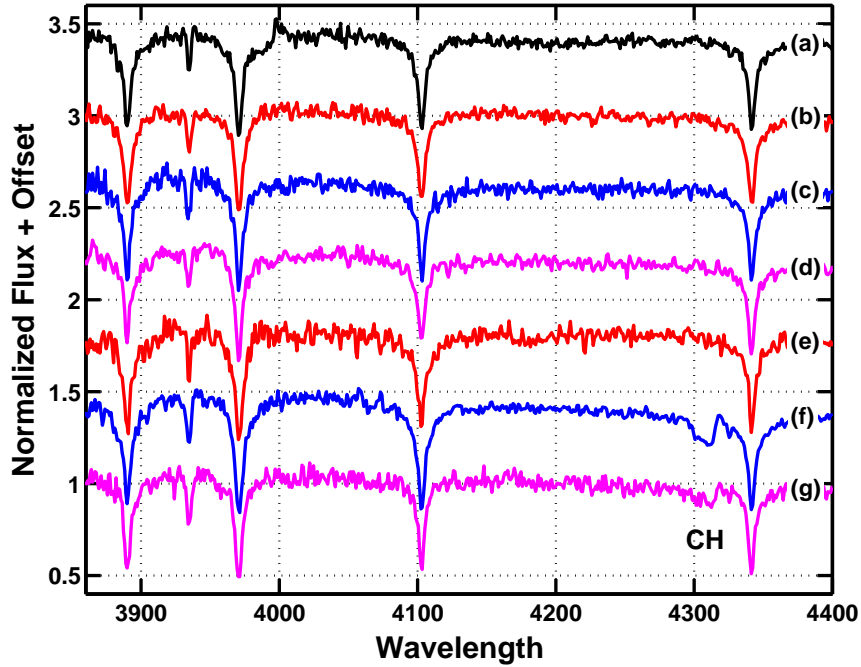}
\caption{ Comparison of a known SDSS EMP with candidate EMP and CEMP stars from Table~\ref{table_likely_emps} that have similar SEDs. The spectra have been normalized to their mean flux in the interval [4150,4270].  The stars shown are: (a)~J130017.20+263238.6, a known SDSS EMP star; (b)~J011844.41+111741.5; (c)~J074748.62+264543.5; (d)~J115315.67+313056.0; (e)~J160257.13+044628.8; (f)~J112147.64-120842.0; and (g)~J080711.80+151458.9.  The last two stars have discernible $^{12}$CH~A-X~(0,0) bands.  \label{fig6_some_emps}}
\end{figure}

\begin{figure}
\epsscale{1.0}
\plotone{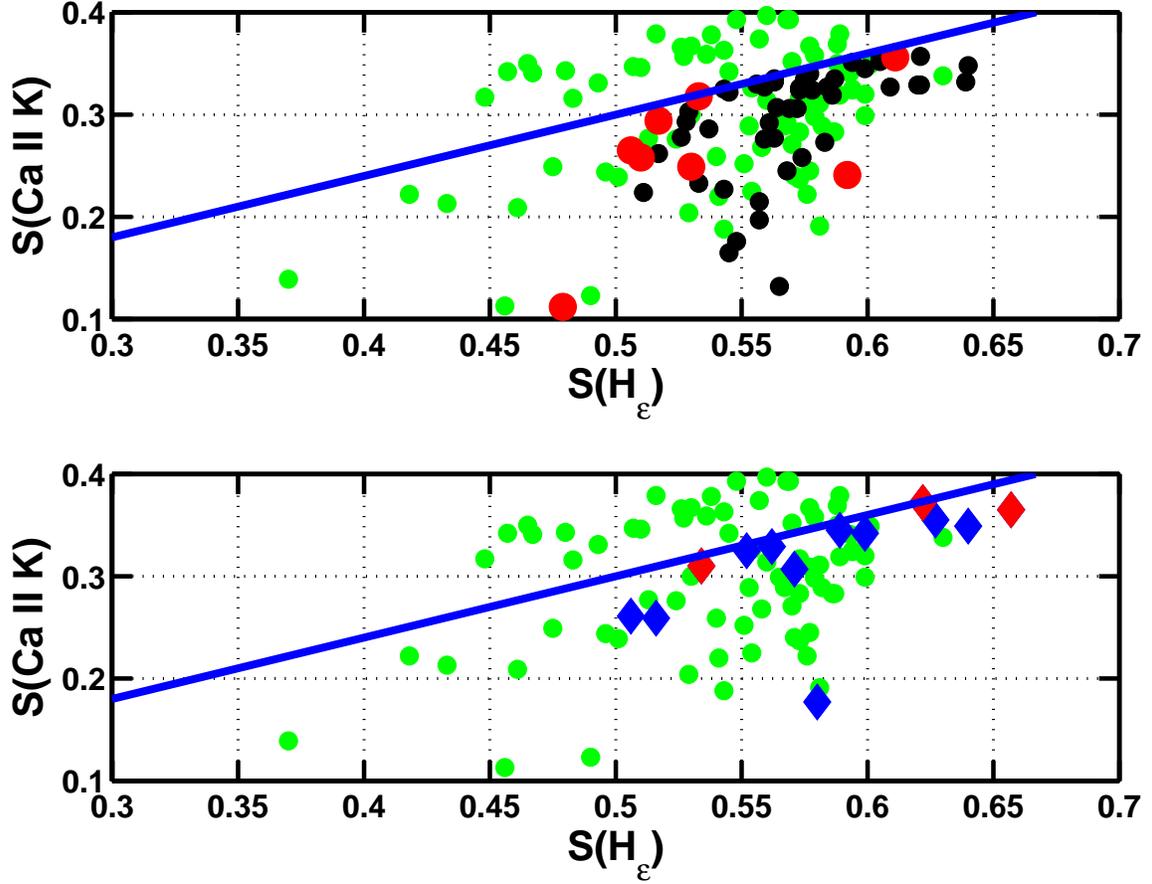}
\caption{Distribution of likely and uncertain candidates.  Note that the scales differ from those in Figure \ref{fig5_HepsCaIIK_partial}.  In the upper plot the stars represented by black dots are the EMP candidates from Table \ref{table_likely_emps} while the stars represented by red dots are the CEMP candidates in that table.  Similarly in the lower plot, the stars represented by blue diamonds are the EMP candidates from Table \ref{table_uncertain_emps} while the stars represented by red diamonds are the CEMP candidates in that table.  For reference, the 
166 known SDSS EMP stars that have both $S(\mathrm{Ca~II~K})$ and $S(\mathrm{H}_\epsilon)$ 
measures are shown in green in both the upper and lower plots.\label{fig7_detailed_distribution_of_candidates}}
\end{figure}

\newpage
\begin{deluxetable}{ccrrrrrrrrrcl}
\tabletypesize{\scriptsize}
\tablecaption{Candidates Likely to be EMP stars\label{table_likely_emps}}
\tablehead{
\colhead{} & \colhead{} & \colhead{} & \colhead{} & \colhead{} & \colhead{$S$} & \colhead{$D$} &  \colhead{$S$} & \colhead{$D$} &  \colhead{$S$} & \colhead{$D$} & \colhead{} & \colhead{}\\
\colhead{Star Name} & \colhead{Plate-Fiber-MJD} & \colhead{$g$} & \colhead{$u$-$g$} & \colhead{$g$-$r$} & \colhead{$\mathrm{Ca~II~K}$} & \colhead{$\mathrm{Ca~II~K}$} &  \colhead{$\mathrm{H}_\epsilon$} & \colhead{$\mathrm{H}_\epsilon$} &  \colhead{$\mathrm{CH~4308}$} & \colhead{$\mathrm{CH~4308}$} & \colhead{SpT} & \colhead{Notes}\\
}

\startdata 
J000401.20$+$154757.6  &  0751-390-52251  &  16.71  &   0.84  &   0.24  &   0.35  &   11.92  &   0.60  &   26.68  &   0.06  &   1.98  &   A0  &     \nodata   \\
J002942.59$-$023723.9  &  4367-62-55566   &  16.70  &   0.84  &   0.25  &   0.28  &   15.21  &   0.53  &   36.30  &   0.05  &   2.92  &   A2II  &     \nodata  \\
J011844.41$+$111741.5  &  4667-34-55868   &  16.98  &   0.84  &   0.25  &   0.23  &   8.52   &   0.53  &   25.14  &   0.02  &   0.64  &   A1III  &     \nodata  \\
J012822.15$+$225013.4  &  5107-428-55940  &  16.28  &   0.84  &   0.23  &   0.27  &   11.89  &   0.51  &   26.33  &   0.24  &   14.65 &   F2III  &   CH  \\
J025052.23$-$032452.0  &  4341-449-55538  &  16.43  &   0.83  &   0.25  &   0.30  &   14.77  &   0.53  &   32.63  &   0.06  &   3.24  &   F0V  &     \nodata  \\
J050110.33$-$033955.1  &  3209-472-54906  &  17.51  &   0.82  &   0.22  &   0.27  &   10.99  &   0.58  &   26.87  &   0.04  &   1.83  &   A0  &     \nodata  \\
J061136.45$+$840554.3  &  2540-509-54110  &  16.31  &   0.79  &   0.25  &   0.29  &   10.49  &   0.56  &   27.39  &   0.05  &   2.16  &   A0  &     \nodata  \\
J073935.14$+$443442.4  &  3663-86-55176   &  16.47  &   0.84  &   0.24  &   0.26  &   14.66  &   0.52  &   37.18  &   0.05  &   3.17  &   F0V  &     \nodata  \\
J074027.05$+$303339.3  &  4448-440-55538  &  17.13  &   0.89  &   0.28  &   0.33  &   13.50  &   0.56  &   28.11  &   0.03  &   1.08  &   B3II  &     \nodata  \\
J074748.62$+$264543.5  &  2055-121-53729  &  17.12  &   0.92  &   0.20  &   0.17  &   4.99   &   0.55  &   22.07  &   0.03  &   1.15  &   A0  &     \nodata  \\
J074943.30$+$670825.1  &  2939-561-54515  &  16.32  &   0.80  &   0.24  &   0.33  &   17.16  &   0.62  &   39.98  &   0.05  &   3.16  &   A0  &     \nodata  \\
J075731.03$+$120716.7  &  2265-239-53674  &  16.87  &   0.86  &   0.21  &   0.33  &   9.93   &   0.57  &   17.96  &   0.06  &   1.96  &   A0  &     \nodata  \\
J080326.35$+$325146.9  &  3756-688-55505  &  17.23  &   0.89  &   0.29  &   0.29  &   11.20  &   0.53  &   24.86  &   0.08  &   3.34  &   A2II  &     \nodata  \\
J080336.58$+$053430.6  &  2056-190-53463  &  16.93  &   0.83  &   0.23  &   0.29  &   8.12   &   0.54  &   18.45  &   0.05  &   1.66  &   A0  &     \nodata  \\
J080711.80$+$151458.9  &  4499-608-55572  &  17.37  &   0.84  &   0.26  &   0.25  &   8.15   &   0.53  &   21.11  &   0.10  &   3.89  &   A2II  &   CH  \\
J081554.26$+$472947.9  &  3693-408-55208  &  16.79  &   0.85  &   0.29  &   0.11  &   3.53   &   0.48  &   19.37  &   0.30  &   14.33 &   F3/F5V  &   CH  \\
J081754.08$+$451035.6  &  3691-30-55274   &  17.55  &   0.85  &   0.24  &   0.29  &   9.25   &   0.52  &   18.30  &   0.16  &   5.90  &   A1III  &   CH  \\
J084932.30$+$583458.3  &  1784-601-54425  &  16.93  &   0.89  &   0.25  &   0.32  &   11.39  &   0.57  &   25.57  &   0.06  &   2.18  &   A0  &     \nodata  \\
J094143.13$+$361028.3  &  3223-170-54865  &  17.04  &   0.89  &   0.20  &   0.33  &   14.11  &   0.61  &   33.81  &   0.05  &   2.33  &   A0  &     \nodata  \\
J094147.00$-$030131.2  &  3782-282-55244  &  16.77  &   0.81  &   0.25  &   0.31  &   14.13  &   0.56  &   33.10  &   0.08  &   4.33  &   A2II  &     \nodata  \\
J094940.20$+$270703.6  &  2342-234-53742  &  16.06  &   0.82  &   0.28  &   0.36  &   13.99  &   0.61  &   32.18  &   0.28  &   13.47 &   A0  &   CH  \\
J101217.52$+$262808.5  &  2386-557-54064  &  16.99  &   0.84  &   0.21  &   0.33  &   13.56  &   0.56  &   29.41  &   0.04  &   1.44  &   A0  &     \nodata  \\
J105613.05$+$451845.1  &  4689-708-55656  &  16.93  &   0.78  &   0.26  &   0.32  &   14.34  &   0.55  &   30.41  &   0.08  &   3.54  &   A1III  &     \nodata  \\
J105928.93$+$470710.3  &  2390-2-54094    &  17.53  &   0.88  &   0.21  &   0.13  &   3.25   &   0.56  &   19.08  &   0.06  &   1.93  &   A0  &     \nodata  \\
J110821.20$-$164648.2  &  2690-165-54211  &  16.38  &   0.90  &   0.21  &   0.36  &   10.11  &   0.62  &   17.31  &   0.08  &   2.70  &   A0  &     \nodata  \\
J112147.64$-$120842.0  &  2859-106-54570  &  16.19  &   0.85  &   0.18  &   0.24  &   10.06  &   0.59  &   27.99  &   0.19  &   10.48 &   A0  &   CH  \\
J113453.71$+$663943.5  &  2858-111-54498  &  17.69  &   0.93  &   0.18  &   0.24  &   6.71   &   0.57  &   19.94  &   0.01  &   0.34  &   A0  &    \tablenotemark{a}   \\
J115315.67$+$313056.0  &  1991-167-53446  &  16.67  &   0.81  &   0.24  &   0.18  &   5.18   &   0.55  &   20.15  &   0.08  &   2.78  &   A0  &     \nodata  \\
J120253.60$-$002620.4  &  2892-194-54552  &  15.77  &   0.81  &   0.20  &   0.26  &   13.61  &   0.57  &   38.13  &   0.03  &   1.36  &   A0  &     \nodata  \\
J121619.51$+$494126.3  &  2919-31-54537   &  17.94  &   0.83  &   0.24  &   0.35  &   7.46   &   0.64  &   16.39  &   0.05  &   1.46  &   A0  &     \tablenotemark{b}   \\
J124643.01$+$484453.1  &  2898-288-54567  &  16.89  &   0.79  &   0.26  &   0.33  &   13.45  &   0.56  &   26.14  &   0.05  &   2.49  &   A0  &     \nodata  \\
J124947.16$+$502202.2  &  2898-468-54567  &  17.33  &   0.84  &   0.20  &   0.36  &   11.56  &   0.61  &   24.99  &   0.07  &   2.58  &   A0  &      \tablenotemark{b}   \\
J125005.10$+$102156.4  &  2963-474-54589  &  16.29  &   0.94  &   0.26  &   0.33  &   15.44  &   0.54  &   34.05  &   0.02  &   0.94  &   A0  &     \nodata  \\
J125203.88$+$500015.3  &  2898-498-54567  &  16.42  &   0.88  &   0.21  &   0.32  &   19.01  &   0.59  &   47.06  &   0.06  &   3.55  &   A0  &     \nodata  \\
J125246.46$+$205516.4  &  5988-46-56072   &  17.24  &   0.77  &   0.25  &   0.22  &   7.01   &   0.51  &   19.99  &   0.04  &   1.67  &   A1III  &     \nodata  \\
J125556.07$+$101715.1  &  3234-355-54885  &  17.67  &   0.84  &   0.22  &   0.32  &   8.41   &   0.53  &   16.21  &   0.25  &   8.85  &   A0  &   CH  \\
J130149.62$+$360023.7  &  2016-350-53799  &  16.36  &   0.89  &   0.20  &   0.33  &   11.68  &   0.57  &   26.87  &   0.07  &   2.73  &   A0  &     \nodata  \\
J130701.00$+$345611.2  &  2016-247-53799  &  16.59  &   0.84  &   0.22  &   0.34  &   10.79  &   0.60  &   21.82  &   0.07  &   2.27  &   A0  &     \nodata  \\
J131116.58$+$001237.6  &  2901-464-54652  &  16.20  &   0.90  &   0.14  &   0.20  &   6.69   &   0.56  &   27.91  &   0.04  &   1.60  &   A0  &    \tablenotemark{c}  \\
J132127.81$+$173341.5  &  2605-474-54484  &  16.35  &   0.85  &   0.19  &   0.35  &   16.94  &   0.59  &   39.16  &   0.05  &   2.23  &   A0  &     \nodata  \\
J133224.76$+$273058.2  &  2245-450-54208  &  16.78  &   0.83  &   0.22  &   0.23  &   7.34   &   0.54  &   23.73  &   0.06  &   2.06  &   A0  &     \nodata  \\
J134003.05$+$095755.8  &  2928-412-54614  &  16.59  &   0.85  &   0.19  &   0.32  &   22.13  &   0.58  &   51.28  &   0.05  &   3.13  &   A0  &     \nodata  \\
J134648.05$+$273145.4  &  2904-136-54574  &  15.97  &   0.86  &   0.20  &   0.31  &   27.14  &   0.57  &   60.19  &   0.05  &   4.17  &   A0  &     \nodata  \\
J134847.62$+$410805.0  &  1377-571-53050  &  16.78  &   0.85  &   0.22  &   0.35  &   11.97  &   0.60  &   25.30  &   0.04  &   1.23  &   A0  &     \nodata  \\
J134913.28$+$173602.7  &  2905-133-54580  &  16.83  &   0.81  &   0.18  &   0.34  &   12.80  &   0.59  &   23.87  &   0.05  &   2.06  &   A0  &     \nodata  \\
J143303.99$+$205157.2  &  2964-306-54632  &  16.52  &   0.85  &   0.18  &   0.33  &   12.78  &   0.57  &   25.63  &   0.05  &   2.04  &   A0  &     \nodata  \\
J143943.92$-$021051.1  &  0919-78-52409   &  16.50  &   0.85  &   0.21  &   0.33  &   10.13  &   0.64  &   26.06  &   0.01  &   0.45  &   A0  &     \nodata  \\
J144811.72$+$384601.9  &  1350-67-52786   &  16.64  &   0.85  &   0.21  &   0.33  &   10.44  &   0.62  &   26.06  &   0.08  &   2.86  &   A0  &     \nodata  \\
J145409.78$-$001111.0  &  3314-206-54970  &  17.31  &   0.87  &   0.18  &   0.26  &   7.31   &   0.51  &   17.89  &   0.20  &   7.77  &   A0  &   CH  \\
J151932.49$+$063610.7  &  2902-104-54629  &  16.48  &   0.85  &   0.17  &   0.34  &   17.42  &   0.58  &   35.08  &   0.04  &   2.13  &   A0  &     \nodata  \\
J154045.84$+$034914.7  &  0594-349-52045  &  17.00  &   0.84  &   0.25  &   0.33  &   9.72   &   0.58  &   21.58  &   0.09  &   3.29  &   A0  &   \tablenotemark{b}  \\
J154811.21$+$261450.9  &  2459-309-54544  &  17.05  &   0.84  &   0.20  &   0.31  &   12.59  &   0.57  &   25.19  &   0.04  &   1.71  &   A0  &     \nodata  \\
J160257.13$+$044628.8  &  2175-294-54612  &  17.27  &   0.89  &   0.20  &   0.28  &   7.32   &   0.56  &   16.89  &   0.08  &   2.66  &   A0  &    \tablenotemark{b}  \\
J160443.30$+$462513.6  &  0813-140-52354  &  16.55  &   0.84  &   0.22  &   0.34  &   11.55  &   0.57  &   21.50  &   0.04  &   1.52  &   A0  &     \nodata  \\
J173239.50$+$640049.1  &  2551-164-54552  &  16.71  &   0.83  &   0.20  &   0.21  &   6.46   &   0.56  &   21.57  &   0.05  &   1.72  &   A0  &     \nodata  \\
J220055.92$+$095253.5  &  5065-476-55739  &  17.87  &   0.79  &   0.26  &   0.28  &   9.74   &   0.56  &   24.41  &   0.05  &   1.88  &   A2II  &     \nodata  \\
J231031.86$+$031847.6  &  4288-862-55501  &  16.84  &   0.80  &   0.25  &   0.34  &   11.20  &   0.56  &   23.95  &   0.01  &   0.22  &   A2II  &    \tablenotemark{b}  \\
\enddata
\tablenotetext{a}{Steepest SED slope of all candidate EMP stars, possible comparison spectra are quite noisy.}
\tablenotetext{b}{Candidate has noisy SDSS spectrum.}
\tablenotetext{c}{Very weak Ca~II~K line.}
\end{deluxetable}

We show in Figure~\ref{fig6_some_emps} some examples from
Table~\ref{table_likely_emps} that were chosen because their SEDs were
similar to the SED of a well-established EMP, J130017.20+263238.6.  Two
independent studies of J130017.20+263238.6 found very similar stellar
parameters ($T_{\mathrm{eff}}$,$\log g$,$[$Fe/H$]$): (6393,~4.00,~-3.65) from
\citet{bonifacio12} and (6450,~4.0,~-3.53) from \citet{aoki13}.  Two of
these stars, J080711.80+151458.9 and J112147.64-120842.0, show evidence
of $^{12}$CH~A-X~(0,0) absorption and therefore are candidate CEMP stars.

\subsubsection{Uncertain EMP candidates and rejected stars} \label{uncertain}

Not all of the 109 stars selected for closer examination fell nicely
within the boundaries of spectral characteristics defined by the 168
known SDSS EMP stars.  Thirteen stars were found to be arguably close
to the spectrum space defined by the 168 known SDSS EMP stars, but
could not convincingly be shown to lie comfortably within that group.
The problem often lay in the fact that the spectrum of the candidate
EMP and/or the spectra of the closest known EMP stars were too noisy
to make a clean judgment. In a number of cases, a single
100~-~300~\AA \ stretch of a particular candidate's SED did not fit by
a wide margin the SEDs of any potentially matching known EMP stars.
We suspect localized poor fluxing by the SDSS pipeline in these
cases. There were also a few stars that had poor SED fits or poor
Balmer line fits with known EMP stars, but showed weak Ca~II~K
absorption and CH absorption.  We classified all 13 stars as
``uncertain'' and show their basic data in
Table~\ref{table_uncertain_emps}. These stars warrant some
consideration since they may be EMP stars, or even CEMP stars.
However, we feel the likelihood is not as high as for those in
Table~\ref{table_likely_emps}.

Before closing this subsection, we wish to note the unexpected SDSS
classification of \\J202109.02+601605.3 as a cataclysmic variable.
While there are stars in the Szkody tabulation (see Section~\ref{cvs})
of CV stars that do have SEDs somewhat like that of
J202109.02+601605.3, they all show prominent emission lines.  The
spectrum of our candidate EMP star appears to have no emission lines
at all.  Disregarding its emission lines, the CV J204448.92-045928.8
is the closest match in the overall SED shape.  However, since its
secondary is estimated to be be K4 or K5 \citep{peters05}, it displays
numerous strong absorption features throughout its SDSS spectrum that are not
apparent in the much more featureless J202109.02+601605.3.  On the
other hand, J202109.02+601605.3's SED and Balmer line widths/depths
are bracketed by two known SDSS EMP stars, J230959.55+230803.1 and
J170339.60+283649.9.  For this reason, we offer J202109.02+601605.3 as
an ``uncertain'' EMP candidate.  In this vein, it bears
  remarking here that two stars in our tabulation of 168 known SDSS
  EMP stars, J082118.18+181931.8 and J092912.33+023817.0, also are
  classified by the SDSS pipeline as being CV stars. Neither star's
  SDSS spectrum shows evidence of emission; although
  J092912.33+023817.0 shows especially weak Ca II K, in keeping with
  its ultra metal-poor status \citep{caffau16}, neither star's
  spectrum is otherwise distinctive.  It is interesting that the SDSS
  spectral classification algorithm identifies these low-metallicity
  objects as CVs.

\begin{deluxetable}{ccrrrrrrrrrcl}
\tabletypesize{\scriptsize}
\tablecaption{Stars with ``Uncertain'' Classification\label{table_uncertain_emps}}
\tablehead{
\colhead{} & \colhead{} & \colhead{} & \colhead{} & \colhead{} & \colhead{$S$} & \colhead{$D$} &  \colhead{$S$} & \colhead{$D$} &  \colhead{$S$} & \colhead{$D$} & \colhead{} & \colhead{} \\
\colhead{Star Name} & \colhead{Plate-Fiber-MJD} & \colhead{$g$} & \colhead{$u$-$g$} & \colhead{$g$-$r$} & \colhead{$\mathrm{Ca~II~K}$} & \colhead{$\mathrm{Ca~II~K}$} &  \colhead{$\mathrm{H}_\epsilon$} & \colhead{$\mathrm{H}_\epsilon$} &  \colhead{$\mathrm{CH~4308}$} & \colhead{$\mathrm{CH~4308}$} & \colhead{SpT} & \colhead{Notes}\\
}

\startdata 
J112736.01$+$592431.3  &  3211-235-54852  &  17.57  &   0.89  &   0.15  &   0.35  &   10.26  &   0.63  &   23.81  &   0.11  &   4.28  &   A0  & \nodata  \\
J121925.37$-$000848.6  &  2558-315-54140  &  16.98  &   0.87  &   0.19  &   0.37  &   15.01  &   0.62  &   30.30  &   0.19  &   8.96  &   A0  &  CH   \\
J130619.63$+$394942.4  &  2900-390-54569  &  16.55  &   0.93  &   0.09  &   0.36  &   13.08  &   0.66  &   31.14  &   0.16  &   6.41  &   A0  &  CH   \\
J133936.19$+$103304.7  &  2903-404-54581  &  14.90  &   0.85  &   0.20  &   0.34  &   17.89  &   0.60  &   46.47  &   0.05  &   2.34  &   A0  & \nodata  \\
J141022.25$+$430151.5  &  6058-980-56090  &  16.78  &   0.84  &   0.26  &   0.31  &   11.88  &   0.53  &   25.30  &   0.15  &   5.83  &   A2II  &  CH   \\
J144501.72$-$010925.4  &  2909-282-54653  &  15.95  &   0.87  &   0.18  &   0.31  &   22.02  &   0.57  &   48.43  &   0.04  &   2.60  &   A0  & \nodata  \\
J152706.80$+$385509.0  &  2911-146-54631  &  17.24  &   0.87  &   0.21  &   0.34  &   15.81  &   0.59  &   24.88  &   0.08  &   3.82  &   A0  & \nodata  \\
J152735.78$+$383933.4  &  2936-97-54626   &  17.91  &   0.87  &   0.18  &   0.18  &   4.59   &   0.58  &   18.08  &   0.08  &   2.77  &   A0p  &  \tablenotemark{a}  \\
J155159.34$+$253900.6  &  2459-256-54544  &  17.11  &   0.91  &   0.27  &   0.26  &   10.30  &   0.52  &   19.85  &   0.07  &   2.97  &   A0  & \nodata  \\
J173632.66$+$643218.9  &  2561-559-54597  &  18.08  &   0.86  &   0.23  &   0.35  &   9.41   &   0.64  &   21.66  &   0.07  &   2.40  &   A0  & \nodata  \\
J202109.02$+$601605.3  &  2564-596-54275  &  16.98  &   0.75  &   0.34  &   0.33  &   8.94   &   0.55  &   18.93  &   0.05  &   1.84  &   CV  & \nodata  \\
J220012.99$+$092501.4  &  4095-8-55497    &  17.48  &   0.80  &   0.23  &   0.26  &   8.86   &   0.51  &   20.48  &   0.10  &   3.86  &   A1III  & \nodata  \\
J235252.78$+$055513.4  &  4405-688-55854  &  17.43  &   0.83  &   0.27  &   0.33  &   10.52  &   0.56  &   22.86  &   0.08  &   2.93  &   A2II  & \nodata  \\
\enddata
\tablenotetext{a}{Markedly weak Ca~II~K line.}
\end{deluxetable}

\subsubsection{Distribution of likely and uncertain candidates} \label{distribution}

We close with a brief examination of the distribution of the stars in
Tables \ref{table_likely_emps} and \ref{table_uncertain_emps} in the
($S(\mathrm{H}_\epsilon)$,$S(\mathrm{Ca~II~K})$) plane. By comparing Figure \ref{fig5_HepsCaIIK_partial} with Figure
\ref{fig7_detailed_distribution_of_candidates}, we see how the stars
chosen for inclusion in those tables are distributed relative to all
the stars selected by the diagonal cut in Figure
\ref{fig5_HepsCaIIK_partial}.  Perhaps the most striking aspect of the
distribution of the chosen candidates is that none of them have
$S(\mathrm{H}_\epsilon)$~$>$~0.66, whereas the distribution of the
selected (blue) stars in Figure \ref{fig5_HepsCaIIK_partial} extends
to $S(\mathrm{H}_\epsilon)$~$=$~0.84.  In fact, the region with
$S(\mathrm{H}_\epsilon)$~$>$~0.66 contains most of the selected stars,
fully 500 of the 695 selected, or 72~\%.  The majority of the stars in
this region were deleted because they were white dwarfs, white
dwarf/cool star binaries, or early-type stars with very broad H Balmer
lines.  Larger values of $S(\mathrm{H}_\epsilon)$ generally go
hand-in-hand with increasing line breadth.  The rest were deleted
because they had defective spectra.  Most were deleted in the coarse
cull described in Section~\ref{culling1}, the rest (18) were deleted
in the fine cull of Section~\ref{culling2}.

The EMP and CEMP candidates from Table \ref{table_likely_emps} shown
in the upper plot of Figure
\ref{fig7_detailed_distribution_of_candidates} appear to follow
the roughly same distribution as the known EMP/CEMP stars.  The
situation seems somewhat different in the lower plot where we see the
distribution for the uncertain EMP/CEMP candidates.  Here there
appears to be a tendency, perhaps statistically significant, for the
candidate stars to be more concentrated near the line defining the
diagonal cut.  We have no ready explanation at present for why the
stars in Table \ref{table_uncertain_emps} would tend to have higher
$S(\mathrm{Ca~II~K})$ at a given $S(\mathrm{H}_\epsilon)$.

\subsection{Application II --- A Direct Route to the CEMP Stars} \label{cemps}

\begin{figure}
\epsscale{0.5}
\plotone{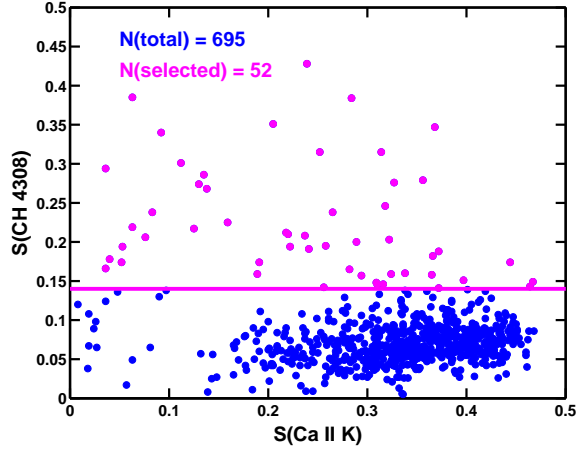}
\caption{Culling potential CEMP candidates.  All the stars in the 
displayed portion of the ($S(\mathrm{Ca~II~K})$,$S(\mathrm{CH~4308})$) plane satisfy 
the constraints (1.0~$\le$~($u$-$z$)~$\le$~1.7), (median~S/N)~$>$~35), 
and ($S(\mathrm{Ca~II~K})$~$<$~0.6~$\times$~$S(\mathrm{H}_\epsilon)$).  The stars shown 
in magenta \emph{additionally} satisfy the constraint that 
($S(\mathrm{CH~4308})$~$>$~0.14), a cut that extracts high-likelihood CEMP 
candidates. The color-coded ``N''s indicate how many stars are in the 
figure area and how many were subselected. \label{fig8_CaIIKCH}}
\end{figure}

Tables~\ref{table_likely_emps} and~\ref{table_uncertain_emps} contain
11 stars with visible CH absorption at SDSS resolution making them
candidate CEMP stars.  While we deduced the presence
of CH by direct inspection of the spectra during the culling phase
described in Section~\ref{culling2}, there is obviously a more direct
route to CEMP stars via an additional LSP step.  We can quickly identify
likely CEMP stars by taking the 695 stars identified as EMP
candidates by the steps in Section~\ref{selection1} and Section~\ref{selection2} and subjecting
them to an additional cut on the depth of the 4308~\AA~A-X~(0,0)
band head of ${^{12}}$CH, $S(\mathrm{CH~4308})$.  Figure~\ref{fig8_CaIIKCH}
shows the ($S(\mathrm{Ca~II~K})$, $S(\mathrm{CH~4308})$) phase space where we have
chosen to make a cut to select only those stars with
$S(\mathrm{CH~4308})$~$>$~0.14.  This cut produces 52 candidate CEMP
stars. Once the 52 spectra were culled for defective spectra, spectra
whose CH ``signal'' was merely depression from the far blue wing
of H$_\gamma$, and spectra from stars that otherwise fell outside the
bounds of known EMP stars, we were left with 10 of the 11 likely CEMP stars indicated in
Tables~\ref{table_likely_emps} and~\ref{table_uncertain_emps}.  The
one exception, J080711.80+151458.9, (see Figure~\ref{fig6_some_emps}) 
was \emph{not} picked up by the cut in ($S(\mathrm{Ca~II~K})$, $S(\mathrm{CH~4308})$)
phase space simply because its $S(\mathrm{CH~4308})$ value (0.08) was too small
to make the cut.  We could have made the cut value smaller but that
would have been at the expense of having to sort through a significantly
larger population of candidate spectra.  We will return to this point in
Section~\ref{improvements}.

\begin{figure}
\epsscale{0.5}
\plotone{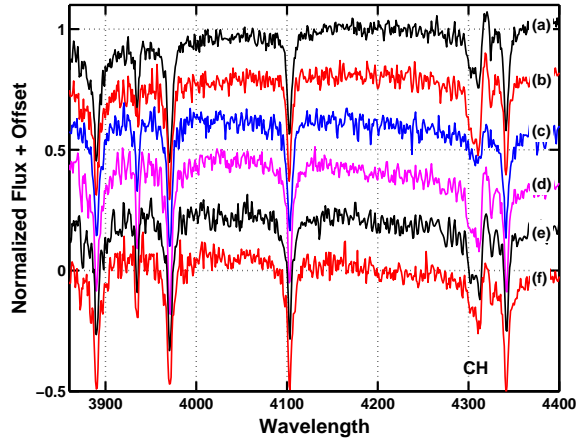}
\caption{The rest of the candidate CEMP stars from Table~\ref{table_likely_emps}. 
The spectra have been normalized to their mean flux in the interval [4150,4270].  
The stars shown are: (a)~J012822.15+225013.4; (b)~J081554.26+472947.9; 
(c)~J081754.08+451035.6; (d)~J094940.20+270703.6; (e)~J125556.07+101715.1; and
(f)~J145409.78-001111.0. The two other CEMP candidates from the table may be 
found in Figure~\ref{fig6_some_emps}. \label{fig9_res_of_cemps}}
\end{figure}

We show in Figure~\ref{fig9_res_of_cemps} the candidate CEMP stars of
Table~\ref{table_likely_emps} that do not already appear in
Figure~\ref{fig6_some_emps}.

\subsection{Application III --- Finding CV stars} \label{cvs}

\begin{figure}
\epsscale{0.5}
\plotone{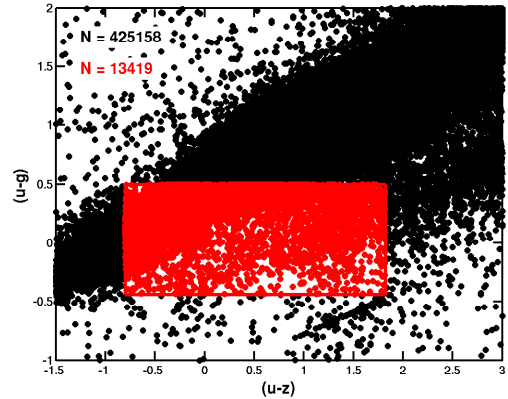}
\caption{First cut in color-color space for CV stars.  All the stars 
present in this portion of phase space are plotted in black.  The 
stars in red were selected using constraints that reflect the 
distribution of Szkody~\emph{et~al}'s SDSS CV stars in our downloaded 
dataset: -0.80~$\le$~($u$-$z$)~$\le$~1.82, and 
-0.44~$\le$~($u$-$g$)~$\le$~0.50. The color-coded ``N''s indicate how 
many stars are in the figure area and how many were subselected. 
\label{fig10_cvs_u-z_u-g}} 
\end{figure}

\begin{figure}
\epsscale{0.5}
\plotone{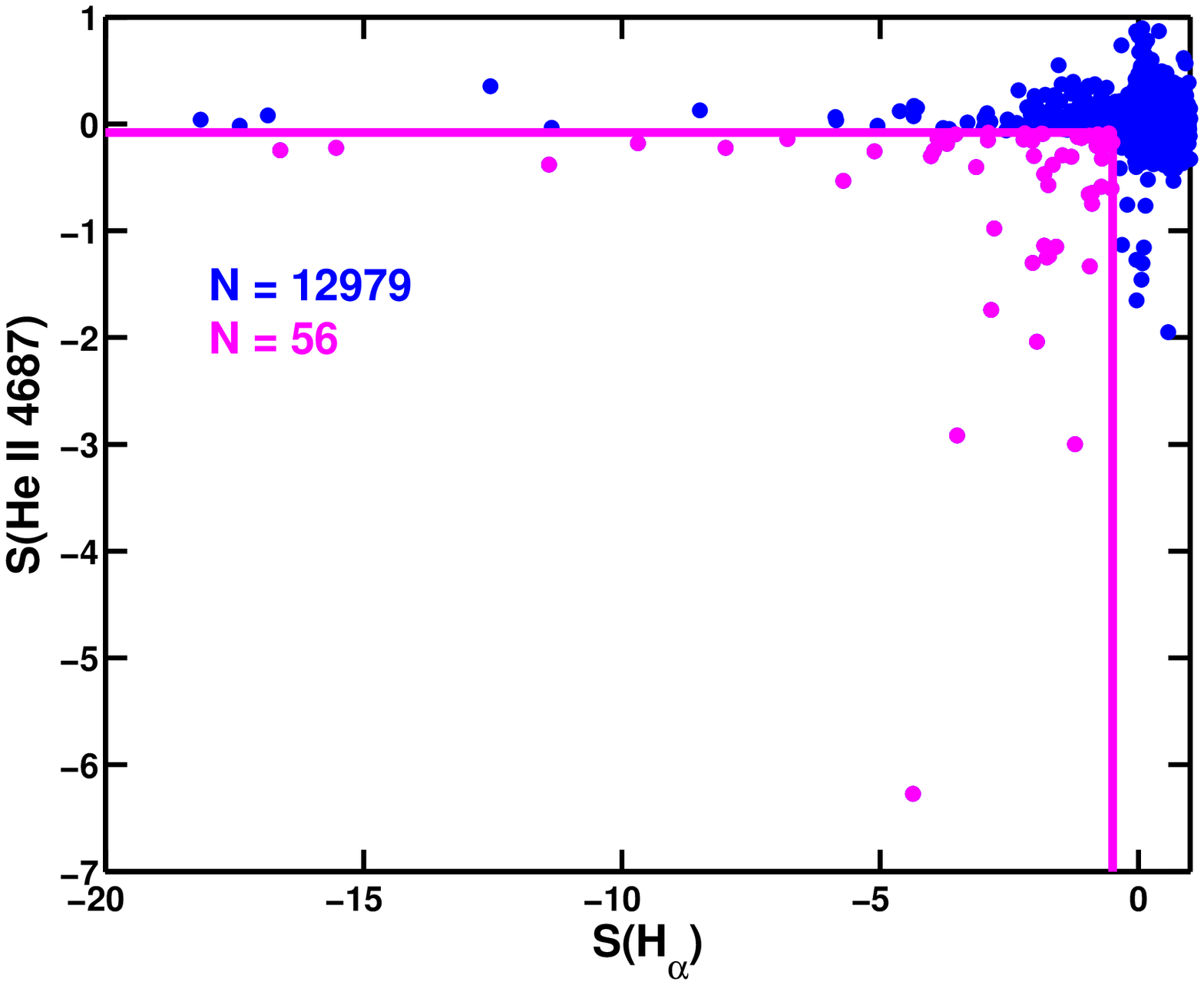}
\caption{ Refining color selections with an additional cut in 
($S(\mathrm{H}_\alpha)$,$S(\mathrm{He~II~4687})$) phase space.  We show here 
primarily the quadrant of this phase space containing stars 
with both H$_\alpha$ \emph{and} He~II~4687 emission ($S(\mathrm{H}_\alpha)$ 
and $S(\mathrm{He~II~4687})$ both negative). The stars marked in blue 
satisfy the cuts we already made in color-color phase spaces: 
$-0.80\le$~($u$-$z$)~$\le$~1.82,~-0.44~$\le$~($u$-$g$)~$\le~$0.50, 
~and~~-0.60~$\le$~($g$-$r$)~$\le 0.90$.  The stars marked in magenta 
\emph{additionally} satisfy the constraints ($S(\mathrm{H}_\alpha)$~$<$~-0.5) 
and $S(\mathrm{He~II~4687})$~$<$~-0.08). The color-coded ``N''s indicate 
the number of stars selected within the figure bounds. 
\label{fig11_cvs_HalphaHeII}}
\end{figure}

Emission line spectra are encountered in a variety of SDSS objects: stars,
galaxies, and QSOs.  In this section we briefly illustrate how the LSP
approach can incorporate emission lines by demonstrating a search for
cataclysmic variable stars. These objects are conveniently available
in our downloaded stellar dataset because they were deliberately
targeted by SDSS investigators~\citep{stoughton02}.  To constrain the
CV star search area, we first made use of the colors of 117 known SDSS
CV stars cataloged by \citet{szkody11}, made available at the
associated WWW site, and present in our downloaded set of SDSS spectra.
Figure~\ref{fig10_cvs_u-z_u-g} shows a part of the ($u$-$z$,$u$-$g$) plane where
we have selected a subregion characteristic of the 117 known
SDSS CV stars.  This cut produced 13,419 candidates out of the 425,158
stars in the plot.  A very slight additional reduction to 13,014
candidates was accomplished by turning to the ($u$-$z$,$g$-$r$) plane
and making another cut informed by the parameters of the Szkody~\emph{et~al}
stars, -0.60~$\le$~($g$-$r$)~$\le$~0.90.

While the two cuts so far have roughly isolated a region occupied by
many CV stars in color space, the selected region is also occupied by
many other stellar types, especially binaries composed of a white
dwarf and an emission line M-dwarf.  To quickly find stars that were
likely to be bona fide CV stars, we adopted the expedient of a
selection based on H and He~II emission line strengths.  While H
emission is an expected feature of CV stars, an examination of the
Szkody~\emph{et~al} stars reveals that He~II emission is by no means
present in all CV stars; see, for example, \citet[][ Table 3]{szkody11}.
However, He~II emission is \emph{also} not present in most of the
stars contaminating the color selected sample.  We therefore
subselected stars with $S(\mathrm{He~II~4687})$~$<$~-0.08 and
$S(\mathrm{H}_\alpha)$~$<$~-0.5 as a crude vehicle to eliminate most of the
contaminating stars.  While using He~II emission this way certainly
must remove some CV stars from our final sample, it does allow us for
illustration to more easily get to a robust subset of He~II emission
CV star candidates.  Our selection in the
($S(\mathrm{H}_\alpha)$,$S(\mathrm{He~II~4687})$) plane is shown in
Figure~\ref{fig11_cvs_HalphaHeII} which shows only stars that have been
selected already by the two previous cuts in color spaces.

Fifty-six stars were selected by the last cut, and after culling stars
with defective spectra, fully 36 remained as apparently good He~II
emission CV star candidates.  Examination of the SDSS spectral type
parameter (subClass) for these 36 candidates showed that all had
been classified CV by the SDSS pipeline, a result that is an
affirmation of our series of cuts in color and emission line spaces.

\begin{figure}
\epsscale{1.0}
\plotone{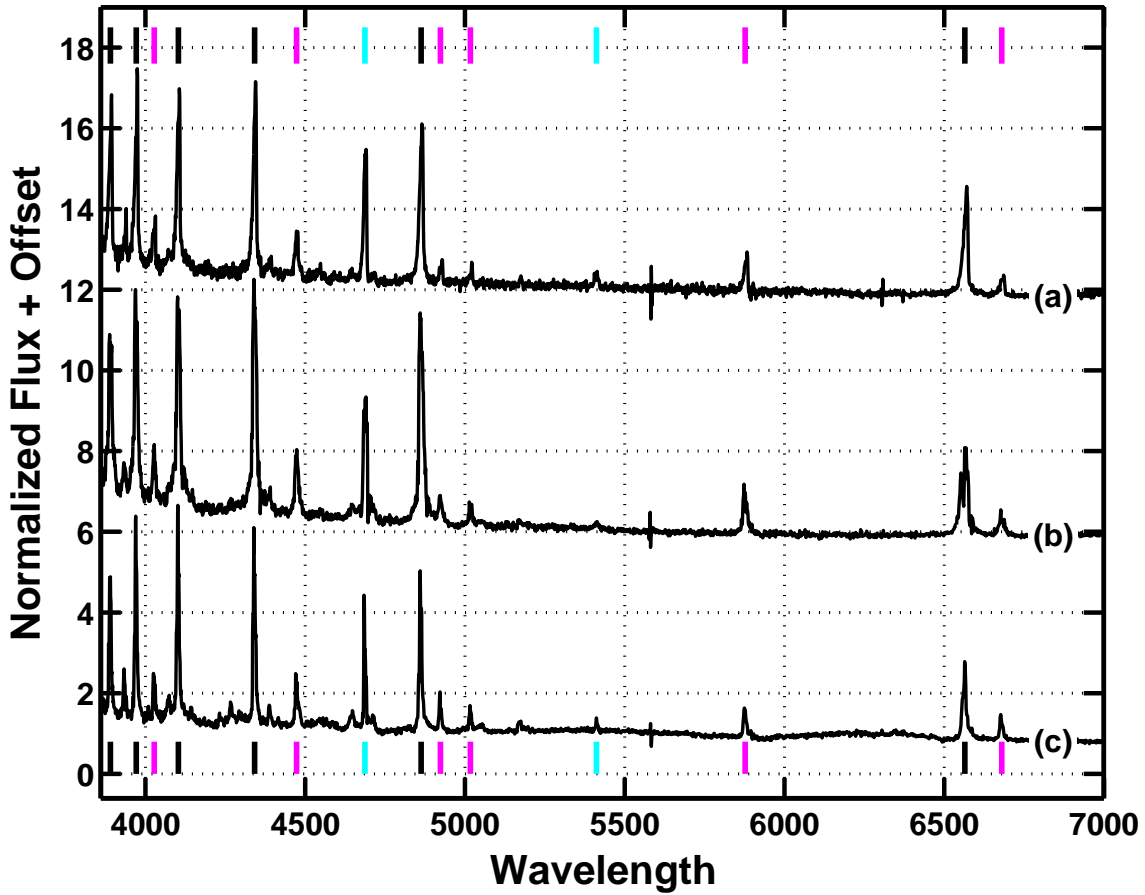}
\caption{Spectra of two new candidate CV stars J014227.07+001729.8~(b) and J153023.63+220646.4~(c) compared with the spectrum of the known SDSS CV star  J100516.61+694136.5~(a).  The wavelengths of the H~I Balmer series are marked at the top and bottom by the vertical black bars; the wavelengths of strong He~I lines are denoted by magenta bars; the two most prominent lines of He~II are marked by cyan bars.    Some He~II transitions are blended at the resolution of this figure with Balmer lines; for example, the He~II 6561.95 is blended with H$_\alpha$.  The flux of each star has been normalized to its mean flux in the interval [5250,6500] and is plotted with an offset. \label{fig12_cvs_2_new}}
\end{figure}

While 32 of the 36 candidates are in the Szkody~\emph{et~al} listing,
4 are not.  Of these, 2 have been discussed in the literature as
CV stars. The first is J131223.49+173659.2, first noted by \citet{vogel08}
based on X-ray variability and then confirmed by them based on its
light curve and spectra as an AM Her type CV star.  The second,
J143209.78+191403.4, has only been mentioned in the literature as a
CV star by \citet{wei13} who recognized it using an interesting data-mining
approach.  They used their Monte Carlo Local Outlier Factor approach
to data-mine over 656,000 SDSS Data Release 8 stellar spectra in
search of spectra that were significant outliers from the norm.  This
interesting approach, which uses principal component analysis to
reduce each spectrum for processing, produced 5 new CV stars from the DR8
dataset, one of which was J143209.78+191403.4.  

It is instructive to understand why our approach did not detect the
other four Wei et al. stars.  One of these stars, J161935.76+524631.8, did not
meet the restrictions of our DR10 download and hence was not considered.
Another of the Wei et al. stars, J111715.91+175741.7 did not have
valid SDSS photometry and so was not considered by us.  Had its
$u$-$z$, $u$-$g$ and $g$-$r$ colors been in the selected ranges, it
would have been detected by us based on its $S(\mathrm{H}_\alpha)$ and
$S(\mathrm{He~II~4687})$ values.  The Wei et al. star
J035747.17-063850.8 survived the selection cuts, but was eliminated in
the culling phase because it did not display convincing evidence of
$\mathrm{He~II~4687}$ emission when its unfortunately low S/N
(median~S/N~=~5.34) spectrum was closely examined.  The fourth star,
J140429.36+172359.6, was eliminated by us because its value of $u$-$g$
(-0.467) missed our cut by 0.027 magnitudes. Even had we chosen a more
negative value for the lower bound on $u$-$g$, it still would have
escaped us because its $S(\mathrm{He~II~4687})$ measurement (+0.041)
fails the cut in that variable.  This provides an interesting
object lesson since the failure in this instance is due to bad
placement of the automatic continuum.  Had the continuum been placed
by eye it would have been lower and the subsequent value of
$S(\mathrm{He~II~4687})$ would have been sufficiently negative
($S(\mathrm{He~II~4687})~\approx$~-0.3) to cause the star to be
selected.  This underlines the importance of good continuum fitting,
especially when cuts are made on comparatively small values of
$\left|S(\lambda_i)\right|$.

The remaining 2 He~II emission CV star candidates of the 36 found,
J014227.07+001729.8 and J153023.63+220646.4, do not appear to be
discussed in the CV star literature according to SIMBAD
\citep{wenger00}.  This may be owing to their rather faint magnitudes.  The
former star does appear without remarks in the catalog of QSOs complied by
\citet{croom09} and was presumably recognized by them as stellar.
According to the SDSS targeting parameter (sourceType),
J014227.07+001729.8 was targeted for SDSS observation as a potential
QSO and J153023.63+220646.4 was targeted as a potential white dwarf.
Additional details for these two stars appear in
Table~\ref{table_new_cvs}.  We show the spectra of these two candidate
He~II emission CV stars in Figure~\ref{fig12_cvs_2_new} along with the
spectrum of J100516.61+694136.5 which is the CV star in our downloaded
Szkody~et~al reference set that most closely matches the two
candidates both in overall spectral distribution and in the relative
intensity of individual emission lines.  (J100516.61+694136.5 is
classified by \citet{wils10} as a dwarf nova and a likely magnetic CV
star.)  Close examination of the three SDSS spectra reveals that they
do differ among themselves in the detailed shape and the wavelength
shifts of peaks of their strong emission lines.

\floattable
\begin{deluxetable}{crrrrcl}
\tabletypesize{\footnotesize}
\tablewidth{0pt}
\tablecaption{Candidate CV stars\label{table_new_cvs}}
\tablehead{
\colhead{Star Name} & \colhead{Plate-Fiber-MJD} & \colhead{$g$} & \colhead{$u$-$g$} & \colhead{$g$-$r$} & \colhead{Spectral Type} & \colhead{Notes}}
\startdata 
J014227.07+001729.8 & 4231-808-55444 & 20.82 &  0.17 &  0.06 & CV & \nodata  \\ 
J153023.63+220646.4 & 3949-814-55650 & 19.08 &  0.05 & -0.09 & CV & \nodata  \\ 
\enddata

\end{deluxetable}

\section{SHORTCOMINGS AND PLANNED IMPROVEMENTS } \label{improvements}

A recurring problem with our approach and quite likely any automated
approach to data-mining spectrum datasets is the necessity to cull
through the results by eye to eliminate those objects that most
definitely do not belong to the desired set of objects, but whose
spectra and/or other measured parameters lead to their automated
capture.  The fault may be due to many factors, often unforeseen,
including flaws in the spectra themselves and inadequate choices of
the constraints used to select the desired objects.  The problem can
be very formidable.  For example, when \citet{rebassa10} used a
$\chi^2$ approach with templates to cull white dwarf-main sequence
binaries from 1.27 million DR6 spectra, they were forced to examine
$\approx$~70,000 spectra by eye to remove interloping QSOs,
galaxies, etc.  With an average of only 5 seconds per spectrum, this
is nearly 100 hours of tedious work.  Even if the automated approach
produces a relatively small number of candidate spectra, the culling
of extraneous objects can be onerous if more than a few seconds must
be spent per spectrum.  As an example, in our
search for CEMP stars (Section~\ref{cemps}), we set the cutoff to be
$S(\mathrm{CH~4308})$~$>$~0.14.  Had we lowered the cutoff to
$S(\mathrm{CH~4308})$~$>$~0.07, we would have captured at least the one CEMP we
had missed from Table~\ref{table_likely_emps} and probably other
similar stars. However, this choice would have produced 356 stars to
be individually examined, nearly 7~$\times$ the previous number.
Since very much more than 5 seconds per spectrum was always required to
decide whether or not a star was an EMP candidate, the
additional effort would not have been trivial.

The researcher is always faced with making a compromise: finding more
candidates for the desired objects versus more time spent
examining individual spectra in order to remove false candidates.  
Our approach, coupling LSPs with the hyperwall, with its
\emph{immediate} visual feedback regarding exactly what is being
selected by the choices made gives users a significant advantage.  
When too many undesirable objects are being selected, one often 
can devise constraints that will eliminate the bulk of these.

In extreme cases, better constraints may require changes to the
spectrum-processing pipeline and the introduction of new spectrum
measures into the mix.  For example, we had adopted linear
interpolation to handle gaps in the SDSS spectra, primarily as a tool
to simplify continuum fitting. This led to false measurements of
$S(\lambda_i)$ and $D(\lambda_i)$ whenever line features fell in
these gaps.  Such erroneous measures were responsible for a sizable
fraction of the false EMP candidates that had to be culled by eye.
Our future pipelines will record ``no-measurement'' when feature
wavelengths fall within a spectrum gap.  In addition, the next
  version of the pipeline will include measurements for several
  widely used line indices including: the Wing ``8c''
  indices \citep{macconnell92}, the Lick indices
  \citep{worthey94,worthey97,trager98}, and molecular indices devised by \citet{brodie86}, \citet{reid95}, and \citep{lepine03}. 

Another improvement that is useful and easy to implement concerns the
wavelengths of molecular band heads.  The current set of band head
wavelengths, drawn as it is from the molecular spectroscopy
literature, is strictly appropriate only to high-resolution spectra.
When the same band systems are viewed at much lower resolutions like
those of the SDSS spectra, the apparent position of maximum band head
absorption shifts, often by a quite noticeable amount.  The effect can
become even more complex because it is also partially dependent on the
strength of the molecular feature, which usually varies greatly with
spectral type.  It appears that a better approach than simply using
the band head wavelengths might be to pick by visual examination
non-band-head wavelengths that are comfortably offset from the band head as it
degrades.  While $\left|S(\lambda_i)\right|$ at these offset positions might not have values
as large as those at the band heads, the overall effect with
regard to monitoring molecular line strength is probably positive.
Selection of the offset wavelengths could be accomplished with
comparatively little effort using appropriate SDSS spectra as a guide.

Finally, it became apparent when sorting through the initial
candidates in Sections~\ref{culling1},~\ref{culling2}, and \ref{cemps}
that measures of strong line width could be very helpful in the
culling process.  Such measures would have greatly simplified the
elimination of hotter stars from the candidate EMP/CEMP lists and
helped excise stars whose H$_\gamma$ wing was sufficient to produce a
false $S(\mathrm{CH~4308})$ signal.  Since line width measures could easily
have other applications as well, it is our goal to add a simple
width measure for strong lines to the next pipeline version.

\section{SUMMARY} \label{summary}

We have described a new approach to the problem of exploring the
contents of very large datasets of spectra.  We showed how the
pairing of linked scatter plots with the visual display capabilities
of NASA's hyperwall creates a powerful and highly flexible spectrum
data exploration tool.  To exploit this tool, we devised a spectrum
reduction pipeline to produce measurements ($S$ and $D$) of atomic and
molecular feature strengths in each individual spectrum of a spectrum
archive.  This pipeline was used to process 569,738 stellar spectra
downloaded from the SDSS DR10 archive, producing nearly 2 billion
individual feature strength measurements to explore with LSPs on the
hyperwall.  The hyperwall allows us to construct and examine
multiple phase-spaces from these feature strength measurements and
other parameters.  The LSP capability on the hyperwall then permits
complex data exploration by allowing the user to subselect objects with
multiple desired characteristics using straightforward graphical
selection tools provided on each hyperwall display.

Using simple selections in just two phase spaces,($u$-$z$,median~S/N) and 
($S(\mathrm{H}_\epsilon)$,$S(\mathrm{Ca~II~K})$), we have revealed 57 new likely 
EMP/CEMP candidates (Table~\ref{table_likely_emps}) and 13 uncertain
EMP/CEMP candidates (Table~\ref{table_uncertain_emps}).  We show that
candidate CEMP stars can be revealed directly, without visually
inspecting spectra, simply by adding one more selection in the
additional ($S(\mathrm{Ca~II~K})$, $S(\mathrm{CH~4308})$) phase space.  Finally, we
demonstrate that CV star candidates with He~II emission
lines can be revealed using simple selections in only three phase
spaces,($u$-$z$,$u$-$g$), ($u$-$z$,$g$-$r$), and ($S(\mathrm{H}_\alpha)$,$S(\mathrm{He~II~4687})$).

The above examples are simple applications of our LSP/hyperwall tool
chosen to illustrate the general principles while using only a small
number of phase spaces.  More complex selection choices involving a
larger number of phase spaces are trivially accomplished.  We believe
that our approach is widely applicable to large spectrum data archives
of stellar, galactic, and QSO spectra, possibly even planetary
spectra, and we are interested in hearing from researchers who may
have suitable search problems.

\vspace{5mm}

\acknowledgments

The authors wish to sincerely thank Karen Huyser, Ruth Peterson, and David Schwenke
for their insightful comments on the draft of this paper.
D.F.C. is indebted to Richard O. Gray for his help in interpreting the
WD/dMe spectra encountered in the search for CV stars.  We also
  wish to thank the anonymous referee for pointing out sections of the
  paper needing additional clarification.  Funding for SDSS-III has
been provided by the Alfred P. Sloan Foundation, the Participating
Institutions, the National Science Foundation, and the U.S. Department
of Energy Office of Science. The SDSS-III web site is
http://www.sdss3.org/.  SDSS-III is managed by the Astrophysical
Research Consortium for the Participating Institutions of the SDSS-III
Collaboration including the University of Arizona, the Brazilian
Participation Group, Brookhaven National Laboratory, Carnegie Mellon
University, University of Florida, the French Participation Group, the
German Participation Group, Harvard University, the Instituto de
Astrofisica de Canarias, the Michigan State/Notre Dame/JINA
Participation Group, Johns Hopkins University, Lawrence Berkeley
National Laboratory, Max Planck Institute for Astrophysics, Max Planck
Institute for Extraterrestrial Physics, New Mexico State University,
New York University, Ohio State University, Pennsylvania State
University, University of Portsmouth, Princeton University, the
Spanish Participation Group, University of Tokyo, University of Utah,
Vanderbilt University, University of Virginia, University of
Washington, and Yale University.  This research has made use of the
SIMBAD database, operated at CDS, Strasbourg, France.
MATLAB\copyright~2015 The MathWorks, Inc. MATLAB and Simulink are
registered trademarks of The MathWorks, Inc. See
www.mathworks.com/trademarks for a list of additional
trademarks. Other product or brand names may be trademarks or
registered trademarks of their respective holders.

\vspace{5mm}

{\itshape Facilities:} \facility{Sloan}

\newpage


\clearpage

\end{document}